\documentclass[%
reprint,
superscriptaddress,
amsmath,amssymb,
aps,
prb
]{revtex4-2}
\usepackage{graphicx}
\usepackage{dcolumn}
\usepackage{bm}
\usepackage{physics}
\usepackage{braket}
\usepackage[colorlinks=true]{hyperref}

\RequirePackage{placeins}
\RequirePackage{listings}
\begin{document}

\let\oldaddcontentsline\addcontentsline
\renewcommand{\addcontentsline}[3]{}

\title{Phonon-number resolution of voltage-biased mechanical oscillators with weakly-anharmonic superconducting circuits}

\author{Mario F. Gely}
\affiliation{%
Kavli Institute of NanoScience, Delft University of Technology,\\
PO Box 5046, 2600 GA, Delft, The Netherlands.
}
\author{Gary A. Steele}

\affiliation{%
Kavli Institute of NanoScience, Delft University of Technology,\\
PO Box 5046, 2600 GA, Delft, The Netherlands.
}%

\date{\today}

\begin{abstract}
Observing quantum phenomena in macroscopic objects, and the potential discovery of a fundamental limit in the applicability of quantum mechanics, has been a central topic of modern experimental physics.
Highly coherent and heavy micro-mechanical oscillators controlled by superconducting circuits are a promising system for this task.
Here, we focus in particular on the electrostatic coupling of motion to a weakly anharmonic circuit, namely the transmon qubit.
In the case of a megahertz mechanical oscillator coupled to a gigahertz transmon, we explain the difficulties in bridging the large electro-mechanical frequency gap.
To remedy this issue, we explore the requirements to reach phonon-number resolution in the resonant coupling of a megahertz transmon and a mechanical oscillator.
\end{abstract}

\maketitle

\noindent 
%
The applicability of quantum phenomena to macroscopic or massive systems has been the topic of intense investigation~\cite{van2008towards,arndt2014testing}, especially in view of the incompatibility between general relativity and quantum mechanics~\cite{karolyhazy1966gravitation,diosi1987favor,penrose1996gravity}.
For example, spontaneous wave function collapse models hypothesize that macroscopicity may provide a fundamental origin to the quantum-to-classical transition~\cite{bassi2013models}.
Micro-mechanical structures oscillating at megahertz frequencies, such as suspended membranes, could be well suited to explore these ideas, due to their large mass and long coherence times~\cite{gely2021superconducting}.
However, the harmonic nature of their motion presents a challenge for their control at a quantum-mechanical level.
%

%
One way this can be addressed is by coupling a non-linear system to the otherwise harmonic oscillator.
One experimental field in which this is done successfully is circuit quantum electrodynamics (QED)~\cite{gu2017microwave}.
In circuit QED, microwave resonators are coupled to superconducting qubits, most often to the weakly-anharmonic transmon qubit, the most prominent building block of a superconducting quantum computer~\cite{koch_charge-insensitive_2007,arute2019quantum}.
If the introduced non-linearity is sufficient to spectrally resolve the number of photons in the resonator, a host of techniques are then available to construct quantum states of the resonator~\cite{hofheinz2008generation,kirchmair2013observation,vlastakis2013,gely2019observation}.
By similarly coupling superconducting circuits to acoustical vibrations through piezoelectricity, phonon number resolution has been achieved, enabling the preparation of quantum states of motion~\cite{o2010quantum,satzinger2018quantum,chu2018creation,arrangoiz2019resolving}.
In these cases however, the high-frequency of the mechanical oscillators make these systems poorly suited to probe macroscopic effects~\cite{gely2021superconducting}.
An alternative, better suited to lower-frequency mechanics, is to embed a voltage-biased oscillator in a circuit~\cite{armour2002entanglement}. 
Coupling superconducting qubits to megahertz mechanical oscillators in this manner has been the topic of multiple experiments, however the coupling was always too small to enable the spectral resolution of phonon-number states~\cite{lahaye2009nanomechanical,pirkkalainen_hybrid_2013,Viennot2018,ma2020nonclassical}.
%

%
%
Here, we explore the reasons behind these difficulties, and explore possible solutions, in the particular case of weakly-anharmonic superconducting circuits.
The results are three-fold.
First, we present a method for analyzing voltage-biased mechanical oscillators embedded in electrical circuits.
More specifically, we show how an equivalent electrical circuit can be derived for the mechanical oscillator, which makes all the tools of circuit quantization available to write a Hamiltonian of the system.
Secondly, we demonstrate the difficulty that lies in obtaining phonon-number resolution of sub-GHz mechanical oscillators coupled to GHz weakly anharmonic circuits.
The large frequency gap causes the break-down of common assumptions in deriving dispersive interactions, and in this case leads to currently unattainable requirements on the transmon coherence time.
Thirdly, we explore the resonant coupling of megahertz mechanical oscillators and superconducting circuits as a solution to the above-mentioned problem.
In particular, we derive the requirement on the transmon and mechanical coherence times to obtain phonon-number resolution of different types of mechanical oscillators.

\section{Coupling mechanism and coupling rate}
\label{sec:1}
\subsection{Intuitive picture of the coupling mechanism}
\begin{figure}
\centering
\includegraphics[width=0.45\textwidth]{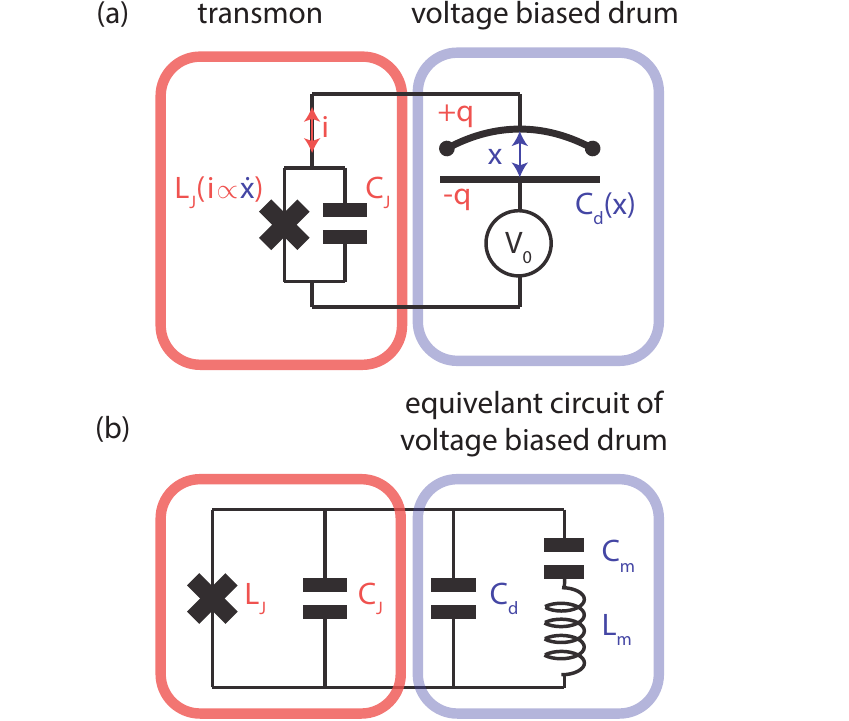}
\caption{
\textbf{Equivalent circuit of a voltage biased mechanical oscillator coupled to a transmon.}
\textbf{(a)} A transmon with Josephson inductance $L_J$ and capacitance $C_J$ is connected to a mechanical oscillator with capacitance $C_d$ biased with a voltage $V_0$.
Movement of the mechanical oscillator will induce a current through the junction, which implements electro-mechanical coupling.
\textbf{(b)} In a circuit equivalent of this system, the voltage biased mechanical oscillator is replaced with a capacitor $C_d$ in parallel with a series $L_m-C_m$ resonator representing the mechanical degree of freedom.
The electrical equivalent allows us to readily apply circuit quantization to derive the system Hamiltonian.
}
\label{fig:drum_and_transmon}
\end{figure}

The coupling mechanism between the mechanical motion and the circuit is illustrated in Fig.~\ref{fig:drum_and_transmon}(a).
The mechanical oscillator plays the role of a mechanically compliant plate of a capacitor $C_d$.
By voltage biasing the capacitor with a voltage $V_0$, a charge $q$ will accumulate on the plates following 
\begin{equation}
	q=C_d(x)V_0,
\end{equation}
where the displacement of the oscillator is denoted by $x$.
Motion will lead to a current flowing in the leads supplying the voltage
\begin{equation}
	i = \dot q = \dot x \frac{\partial C_d}{\partial x} V_0\ .
\end{equation}
If this current flows through a Josephson junction, it will modify the junctions effective inductance.
Specifically, the effective junction inductance will acquire different values for each Fock state of the mechanical oscillators motion.
If the junction is embedded in a resonant circuit (e.g. a transmon qubit) featuring a narrow linewidth, this effect could reveal a spectrum of different peaks corresponding to different Fock states of the mechanical oscillators motion~\cite{schuster_resolving_2007}.
We call this effect phonon-number resolution.

\subsection{Equivalent circuit}
To quantify this effect, we make use of an equivalent circuit for the voltage-biased mechanical oscillator derived in Sec.~\ref{sec:equiv_circuit_drum}, and shown in Fig.~\ref{fig:drum_and_transmon}(b).
Here $C_d$ corresponds to the capacitance formed by the mechanical oscillator (including its static displacement induced by the voltage).
The series composition of $L_m$ and $C_m$ represents the mechanical degree of freedom, and this part of the circuit resonates at the mechanical frequency
\begin{equation}
 	\omega^{V_0}_m = \frac{1}{\sqrt{L_mC_m}} = \sqrt{\frac{k_\text{eff}(V_0)}{m}}\ ,
 \end{equation}
and has an impedence
\begin{equation}
	{Z_m = \sqrt{\frac{L_m}{C_m}} = \frac{D(V_0)^2}{V_0^2C_d^2}\sqrt{k_\text{eff}(V_0)m}}\ ,
\end{equation}
where $m$ is the mass of the mechanical oscillator, $D(V_0)$ corresponds to the distance separating the two capacitive plates of $C_d$, and the effective spring constant of the oscillator $k_\text{eff}(V_0)$ is
\begin{equation}
	k_\text{eff}(V_0) = k-\frac{V_0^2C_d}{D(V_0)^2}\ ,
	\label{eq:k_eff}
\end{equation}
with $k$ the spring constant of the unbiased oscillator.
We refer to the change in spring constant with voltage as electrostatic spring softening.
The condition $k_\text{eff}(V_0)>0$ dictates the maximum applicable voltage, as the mechanical oscillator will become unstable for $k_\text{eff}(V_0)<0$.
\subsection{Hamiltonian of a transmon coupled to a mechanical oscillator}
We consider the voltage biased mechanical oscillator coupled to a transmon qubit composed of a junction $L_J$ and a capacitance $C_J$, such that the total capacitance in parallel of the junction is $C_t = C_J+C_d$ (see Fig.~\ref{fig:drum_and_transmon}(b)).
Following standard circuit quantization techniques~\cite{vool2017introduction}, the Hamiltonian is of the form
\begin{equation}
\begin{split}
  \hat{H} &= \hbar\omega_t' \hat{a}^\dagger\hat{a}-\frac{A}{12} \left(\hat{a}+\hat{a}^\dagger\right)^4\\
  &+\hbar\omega_m' \hat{c}^\dagger\hat{c}-\hbar g \left(\hat{a}-\hat{a}^\dagger\right)\left(\hat{c}-\hat{c}^\dagger\right)\ ,
  \label{eq:rabi_hamiltonian}
\end{split}
\end{equation}
as derived in Sec.~\ref{sec:hamiltonian_derivation}.
Here $\hat{a}$ and $\hat{c}$ are annihilation operators for the transmon and the mechanical oscillator respectively.
The frequency $\omega_t' = 1/\sqrt{L_JC_t}$ is related to the frequency of the first transition ($\ket{g}\leftrightarrow\ket{e}$) of the transmon $\omega_t$ through $\omega_t = \omega_t'-A/\hbar$.
The charging energy $A = e^2/2C_t$ quantifies the anharmonicity of the transmon, and we have made the approximation $A/\hbar\omega_t\le1/20$ in order to write the anharmonicity of the junction as the fourth power of $(\hat{a}+\hat{a}^\dagger)$~\cite{koch_charge-insensitive_2007}.
The mechanical frequency is renormalized when quantizing the circuit $\omega_m' = 1/\sqrt{L_m C_m}\sqrt{(C_m+C_t)/C_t}$. 
The quantity which is most relevant to this discussion is the coupling
\begin{equation}
		g = \frac{\sqrt{\omega_m'\omega_t'}}{2}\sqrt{\frac{1-\left(\omega_m^{V_0}/\omega_m^0\right)^2}{1+\frac{C_J}{C_d}\left(\omega_m^{V_0}/\omega_m^0\right)^2}}\ .\\
\end{equation}
%
The coupling can be increased by having the mechanical oscillator capacitance dominate over the other capacitance of the transmon $C_d\gg C_J$.
Also, increasing the applied voltage lowers the effective spring constant $k_\text{eff}$, which lowers the ratio $\omega_m^{V_0}/\omega_m^0$ and increases the coupling.
This route to increasing the coupling is ultimately limited by the fact that the effective spring constant should remain positive.

\begin{figure*}
\centering
\includegraphics[width=0.9\textwidth]{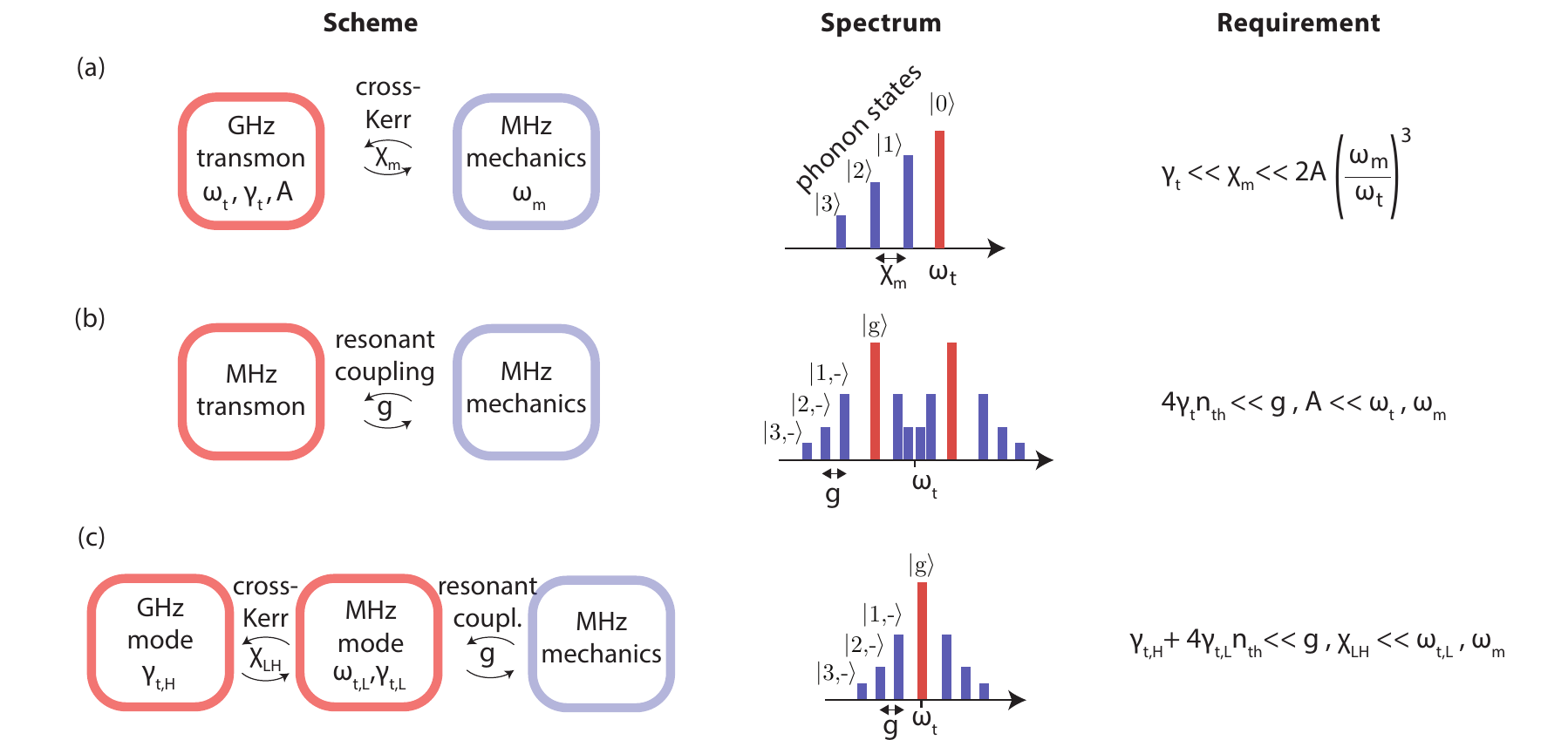}
\caption{
\textbf{Overview of coupling schemes and their requirements for phonon-number resolution.}
The height of the spectral features schematically displayed depends on the occupation of the annotated state.
We denote mechanical Fock states by $\ket{0}$, $\ket{1}$, ... and Jaynes-Cummings~\cite{jaynes_comparison_1963} eigenstates by $\ket{n,-} = \left(\ket{n,g}-\ket{n-1,e}\right)/\sqrt{2}$, where $n$ refers to the number of phonons in the mechanical oscillator, and $g,e$ refers to the ground and first excited state of the transmon.
\textbf{(a)} MHz mechanics coupled to a GHz transmon.
In this case, phonon-number resolution is achieved when the cross-Kerr interaction $\chi_m$ is larger than the transmon line-width $\gamma_t$.
\textbf{(b)} 
Mechanics resonantly coupled to a transmon.
The separation between phonon-dependent spectral peaks is given by the smallest energy scale: the coupling rate $g$ or the transmon anharmonicity $A$.
Here we schematically show the case where $g$ is the smallest, leading to a Jaynes-Cummings spectrum.
Phonon-resolution then relies on these energy scales being larger than the spectral line-width, given by $4\gamma_t n_\text{th}$ if the transmon has the dominating decay rate, where $n_\text{th}$ refers to the thermal occupation of the transmon.
\textbf{(c)} 
A higher, GHz electrical mode is added to the previous setup.
Phonon-number resolution in the high-frequency spectrum comes when the representative linewidth $\gamma_{t,H} + 4\gamma_{t,L} n_\text{th}$, is lower than the coupling rate $g$ and cross-Kerr coupling $\chi_{LH}$.
The spectrum shown here corresponds to the case where the coupling is smaller than the cross-Kerr coupling.
}
\label{fig:requirements}
\end{figure*}

\section{GHz transmon -- MHz mechanical oscillator}
\label{sec:2}
We first explore the conditions to obtain phonon-number resolution in the spectrum of a GHz transmon coupled to a MHz mechanical oscillator as schematically summarized in Fig.~\ref{fig:requirements}(a).

\subsection{Normal-mode picture of the dispersive interaction}

If the coupling is small relative to the frequency difference between the oscillators $g\ll\Delta = \omega_t'-\omega_m'$, the system is better described by transforming the Hamiltonian through a Bogoliubov transformation to~\cite{gely2017nature}
\begin{equation}
\begin{split}
  \hat{H} &\simeq \hbar\tilde\omega_t' \tilde a^\dagger\tilde a-\frac{\tilde A}{2} \tilde a^\dagger\tilde a^\dagger\tilde a\tilde a\\
  &+\hbar\tilde\omega_m' \tilde c^\dagger\tilde c-\frac{\tilde A_m}{2} \tilde c^\dagger\tilde c^\dagger\tilde c\tilde c-\chi_m \tilde a^\dagger\tilde a\tilde c^\dagger\tilde c\ .
\end{split}
  \label{eq:bogo_hamiltonian}
\end{equation}
Here $\tilde a$ and $\tilde c$ are annihilation operators for the normal modes of the system.
Since these can be expressed as a linear combination of the original annihilation and creation operators (without a tilde), they are both electrical and mechanical in nature.
However, due to the assumption $g\ll\Delta$, $\tilde a$ still refers to majoritarily electrical oscillations, and $\tilde c$ majoritarily mechanical oscillations.
Likewise, the frequencies $\tilde\omega_t'$, $\tilde\omega_m'$ and anharmonicity $\tilde A$ acquire only small shifts with respect to their original values (without a tilde).
The detuning between mechanical and electrical frequency is so large in this case, that the sum of oscillation frequencies $\Sigma = \omega_t'+\omega_m'$ is comparable to the difference $\Delta\sim\Sigma\sim\omega_t'$.
This regime forbids the usual rotating-wave approximation when performing the Bogoliubov transformation~\cite{gely2017nature}.
This results in a cross-Kerr interaction between the modes given by
\begin{equation}
	\chi_m \simeq 8 A g^2\frac{ \omega_m'^2}{\omega_t'^4}\ .
	\label{eq:chi}
\end{equation}
This interaction could give rise to phonon-number resolution: if we combine the energy of the electrical mode and the cross-Kerr interaction into one term, $\left(\hbar\tilde\omega_t'-\chi_m\tilde c^\dagger\tilde c\right) \tilde a^\dagger\tilde a$, the frequency of the electrical mode depends on the number of phonons in the mechanical mode $\tilde c^\dagger\tilde c$.
In order for phonon-number resolution to be observable however, the shift per phonon should be greater than the line-width of the electrical mode $\chi_m\gg \gamma_t$.

\subsection{Requirements for phonon-number resolution}
This condition is however difficult to meet given the expression for the cross-Kerr interaction of Eq.~(\ref{eq:chi}).
Indeed, $\chi_m$ is weighted by the anharmonicity $A$, but, to remain in the transmon limit, $A$ has an upper limit $A<\hbar\omega_t'/20$.
$\chi_m$ is also weighed by the coupling squared $g^2$, which also has an upper limit in order to keep the effective spring constant of the mechanical oscillator positive (see Eq.~(\ref{eq:k_eff})).
Taking these facts into account, the cross-Kerr coupling, a function of bias voltage $V_0$, exhibits a maximum value, derived in Sec.~\ref{sec:max_chi} to be
\begin{equation}
	\chi_m<\text{Max}[\chi_m]=\frac{\hbar\omega_t'}{10}\left(\frac{\omega_m^0}{\omega_t'}\right)^3\ .
	\label{eq:GHz_transmon_requirement}
\end{equation}
So if we require a certain $\chi_m$ to resolve phonon-number resolution, based on a transmon linewidth $\gamma_t$, we can determine a minimum value for $\omega_m^0$.
And we can say with certainty that for mechanical oscillators of a lower frequency, it will be mathematically impossible to obtain the required cross-Kerr coupling.
We consider a typical transmon of frequency $\omega_t = 2\pi\times6$ GHz, and dephasing rate $T_2=50\ \mu$s~\cite{kjaergaard2019superconducting}, corresponding to a linewidth $\gamma_t = 2\pi\times3$ kHz.
If we require the cross-Kerr shift to be ten times larger than the linewidth $\chi_m = h\times30$ kHz, we obtain $\omega_m^0>2\pi\times220$ MHz.
So with a typical transmon, the lowest possible mechanical frequency which would enable phonon-number resolution is 220 MHz.
This is however a theoretical optimum, when the bias voltage is such that $k_\text{eff}=0$ (resulting in a 0 frequency mechanical oscillator), so realistically phonon-number resolution is only possible with mechanical oscillators of a much larger frequency.

An alternative approach to phonon-number resolution would be to obtain an anharmonicity in the mechanical mode
$\tilde A_m = \chi^2/4\tilde A$~\cite{gely2017nature} which exceeds the mechanical linewidth.
However, this approach is limited by the influence of the transmon dissipation on the mechanical linewidth, resulting on even more stringent requirements for the transmon coherence time (see Sec.~\ref{sec:Am_greater_than_gammam}).

\section{MHz transmon -- MHz mechanical oscillator}
\label{sec:3}
\subsection{Thermal strong coupling requirement}

Given the difficulties in achieving phonon-number resolution in this dispersive regime, we now study the opposite regime, where the transmon and mechanical oscillator are on resonance.
This would involve a transmon of a low ($\sim$MHz) frequency, and thermally populated even when cooled in a dilution refrigerator.
As shown in Ref.~\cite{gely2019observation}, this does not necessarily mean that one loses access to the quantum nature of the system.
Note that the intermediate regime ($g\ll\Delta\ll \Sigma$) is studied in Sec.~\ref{sec:RWA_regime} and is less favorable than resonant coupling.
As derived in Sec.~\ref{sec:requirements_details_2body}, phonon-number resolution (schematically shown in Fig.~\ref{fig:requirements}(b)) is attained when
\begin{equation}
	4\gamma_t n_\text{th} \ll g, A/\hbar\ ,
		\label{eq:main_ineq_2body}
\end{equation}
where $n_\text{th}$ is the average number of excitations in the transmon due to its thermalization with the environment.
We have defined phonon-number resolution as being able to spectroscopically resolve the $\ket{0}$ and $\ket{1}$ mechanical Fock states.
Since $4\gamma_t n_\text{th}$ is the representative linewidth of the coupled system, broadened by thermal effects, we will refer to the first part of the requirement $4\gamma_t n_\text{th} \ll g$ as thermal strong coupling.
Note that we have assumed that the limiting dissipation rate is that of the electrical degree of freedom.
We also considered adding a second, GHz electrical mode in the spirit of Ref.~\cite{gely2019observation}.
The low-frequency electrical mode and mechanical mode are still coupled resonantly, whilst the two electrical modes are coupled through a cross-Kerr interaction.
The aim of this addition is to observe phonon-number resolution in the spectrum of the high-frequency mode, which is unaffected by thermal fluctuations.
The Hamiltonian describing this system is
\begin{equation}
\begin{split}
	\hat H &= \hbar \omega_{t,H}' \hat a ^\dagger \hat a - A_H\hat a ^\dagger\hat a ^\dagger \hat a \hat a \\
	&+ \hbar \omega_{t,L}' \hat b ^\dagger \hat b - A_L\hat b ^\dagger\hat b ^\dagger \hat b \hat b + \hbar \omega_m'\hat c ^\dagger \hat c\\
	&- \hbar g (\hat b-\hat b^\dagger )(\hat c  -\hat c^\dagger )- \chi_{LH} \hat a ^\dagger\hat a \hat b ^\dagger \hat b\ ,\\
\label{eq:two-mode-hamiltonian}
\end{split}
\end{equation}
where $\hat a$, $\hat b$ and $\hat c$, and the subscripts $H$, $L$ and $m$ correspond to the high frequency (HF) electrical mode, the low-frequency (LF) mode, and the mechanical oscillator respectively.
As derived in Sec.~\ref{sec:requirements_details_3body}, the condition to discriminate different phonon-states in the spectrum of the \textit{high-frequency} mode, schematically shown in Fig.~\ref{fig:requirements}(c), is
\begin{equation}
		\gamma_{t,H}+4\gamma_{t,L} n_{\text{th}}\ll\chi_{LH}/\hbar,g\ll\omega_m,\omega_{t,L}\ ,
		\label{eq:main_ineq_3body}
\end{equation}
where $\gamma_{t,H}$ is the linewidth of the HF mode, and $n_{\text{th}}\gamma_{t,L}$ the thermalization rate of the LF mode.
This setup thus yields a similar requirement (thermal strong coupling) as the case without the high-frequency mode.

\subsection{Quality factor needed and discussion}

\begin{figure}[t!]
\centering
\includegraphics[width=0.45\textwidth]{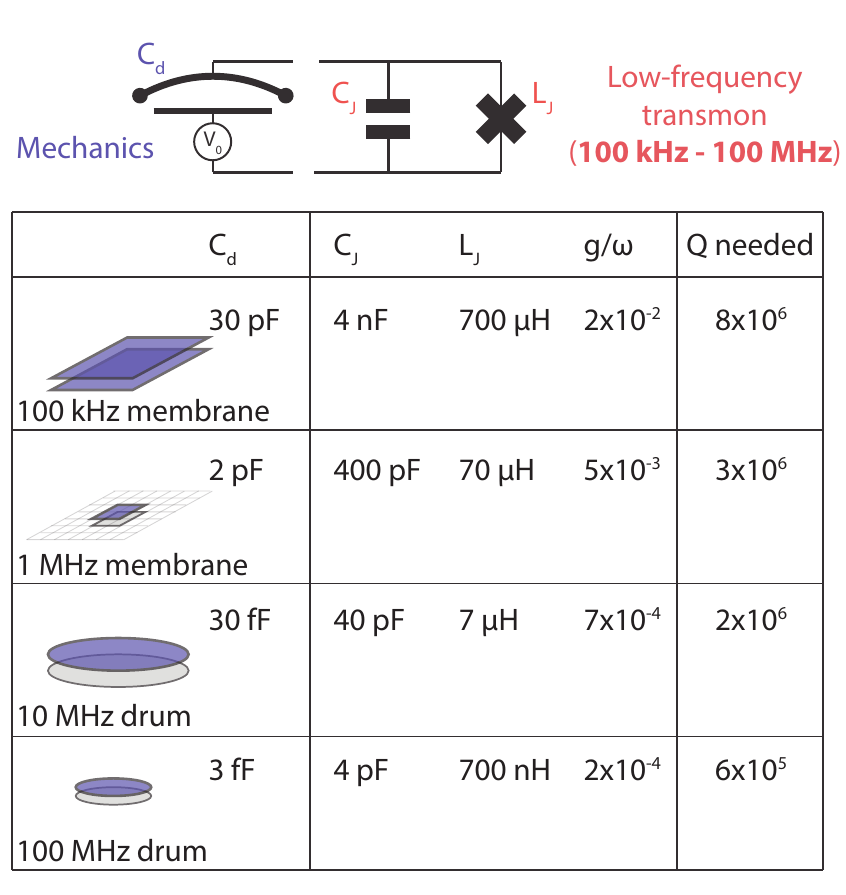}
\caption{
\textbf{Quality factor required for the resonant coupling of a mechanical oscillator and transmon} (schemes (b)-(c) of Fig.~\ref{fig:requirements}).
In the first column we provide a schematic of the mechanical oscillator and the capacitance it implements considering the smallest experimentally achieved gap between the mechanical oscillator and its corresponding electrode.
In the second column, we provide the capacitance and inductance which ensures resonance with the mechanical oscillator and maximum coupling, whilst remaining in the transmon regime.
We then provide the coupling relative to the frequency, assuming that the bias voltage applied to the mechanical oscillator only reduces its frequency through electrostatic spring softening by 10 percent.
Finally we provide the quality factor of both mechanical oscillator and transmon needed for the line-width to be ten times smaller than the coupling (thermal strong coupling).
}
\label{fig:resonant_couplings}
\end{figure}

The technical barrier to phonon-number resolution in the resonant regime, with or without a high-frequency mode, is achieving thermal strong coupling $4\gamma_{t,L} n_{\text{th}}\ll g$.
We calculate in the table of Fig.~\ref{fig:resonant_couplings} the quality factor of the electrical mode resonant with the mechanics necessary to achieve a coupling equal to ten times the relevant linewidth $g = 10\times4\gamma_{t,L} n_{\text{th}}$, for a variety of mechanical oscillators.
This quality factor is given by $Q = \omega_{t}'/\gamma_{t}$ and $Q = \omega_{t,L}'/\gamma_{t,L}$ for a mechanical oscillator coupled to a single transmon or two electrical modes respectively.
Note that the mechanical quality factor whould also match or exceed the electrical quality factor.
We have assumed that the applied voltage brings the oscillator on the cusp of instability $\omega_m^{V_0}/\omega_m^0=0.9$.
Also, we chose capacitances and Josephson inductances which maximize the coupling, whilst maintaining the transmon limit $A/\hbar\omega_t < 1/20$.
The thermal occupancy $n_\text{th}$ is computed with a $20$ milliKelvin temperature.
%
%

The studied mechanical oscillators are 100 kHz~\cite{yuan2015silicon} and 1 MHz~\cite{chen2020entanglement} membranes, which have an area of $A = (1 \text{mm})^2$ and $A = (250 \mu\text{m})^2$ respectively and can be suspended 300 nm above another capacitive plate~\cite{noguchi2016ground}.
Additionally, we consider 10 MHz, $15\mu$m diameter and 100 nm thick membranes suspended 50 nm above an electrode~\cite{teufel2011circuit}, and hypothetical smaller membranes with a $100$ MHz frequency achieved by reducing the area by a factor 10~\cite{rao2006vibration}.
Concerning even higher frequency oscillators, note that gigahertz voltage-biased mechanical oscillators have succesfully been resonantly coupled to transmons~\cite{rouxinol2016measurements}.

For megahertz oscillators however, very large capacitances are required for the electrical mode to reach low frequencies.
A natural solution is to make use of parallel plate capacitors.
However, these typically result in relatively low quality factors due to dielectric losses~\cite{o2008microwave}.
A 170 MHz circuit making use of amorphous Silicon as a dielectric at around 10 milliKelvin was measured~\cite{gely2019observation} to have a quality factor $3\times 10^3$.
Whereas dielectric losses can be mitigated in GHz circuits by constructing capacitors on a single plane, with electric fields traversing extremely clean crystalline substrates and vacuum, this approach may be challenging for the large capacitances needed here.
Obtaining both large capacitors in conjunction with low losses is thus the main roadblock to realizing phonon-number resolution with resonantly coupled mechanical oscillators and transmons.
A few other challenges can be foreseen.
Lowering the transmon frequency translates to very large Josephson inductances.
The values in Fig.~\ref{fig:resonant_couplings} for the Josephson inductance are in the range 700 nH - 700 $\mu$H, corresponding to $h\times$ 230 MHz - 230 kHz in terms of Josephson energy, and 470 pA - 470 fA in terms of cricical current.
Such junctions may be complicated to fabricate or operate.
Additionally, thermally excited current may on average exceed the critical current of the junction, which could make the operation of the device difficult.
One may have to apply sideband cooling to the low-frequency electrical mode ~\cite{gely2019observation}, or the mechanical oscillator to which is it coupled~\cite{teufel2011sideband}.

\section*{Conclusion}

In conclusion, we have shown that phonon-number resolution of MHz mechanical oscillators is not achievable with state-of-the art GHz transmons.
As an alternative, we have proposed the resonant coupling of mechanical motion to a MHz transmon, with and without an additional GHz electrical mode.
The most prominent technical challenge associated with this approach is to construct a low-frequency transmon with a quality factor on par with GHz transmons.
As an outlook, we acknowledge alternative techniques to access the quantum nature of low-frequency mechanical oscillators using superconducting circuits. 
One could increase the anharmonicity of the system, such that the expression for $\chi_m$ is favorably modified by the irrelevance of higher levels of the superconducting circuit.
Recent successes in this direction have been demonstrated using a Cooper-pair box~\cite{Viennot2018,ma2020nonclassical}.
The Fluxonium qubit~\cite{Manucharyan113} also seems like an attractive option at first: it can operate at megahertz operating frequency where it features a large anharmonicity.
However, as its frequency is decreased through flux-biasing, its electrical dipole moment -- which scales the coupling to a voltage biased membrane -- is exponentially suppressed~\cite{PhysRevLett.120.150503}.
Alternatively, one could make use of optomechanical coupling to transfer quantum states in the GHz regime to the lower frequencies of the mechanical oscillation~\cite{reed2017faithful}.
Recent developments regarding coupling through superconducting interference also show promise~\cite{rodrigues2019coupling,kounalakis2019synthesizing,kounalakis2020flux}.

\vspace{5mm}
\textbf{Code availability:} the code used to generate the table of Fig.~\ref{fig:resonant_couplings}, as well as the simulations in the supplementary information is available in Zenodo with the DOI identifier 10.5281/zenodo.4292367 .

\vspace{5mm}
\textbf{Acknowledgments:}
the authors acknowledge funding from the European Union’s Horizon 2020 research and innovation programme under grant agreement 681476 - QOM3D.

\FloatBarrier
\clearpage
\onecolumngrid
\begin{center}
{\huge \textbf{Supplementary information}}
\end{center}
\makeatletter
   \renewcommand\l@section{\@dottedtocline{2}{1.5em}{1em}}
   \renewcommand\l@subsection{\@dottedtocline{2}{3.5em}{1em}}
   \renewcommand\l@subsubsection{\@dottedtocline{2}{5.5em}{1em}}
\makeatother

\let\addcontentsline\oldaddcontentsline


\renewcommand{\thesection}{\arabic{section}} 

\onecolumngrid

\let\oldaddcontentsline\addcontentsline
\renewcommand{\addcontentsline}[3]{}
\let\addcontentsline\oldaddcontentsline


\renewcommand{\theequation}{S\arabic{equation}}
\renewcommand{\thefigure}{S\arabic{figure}}
\renewcommand{\thetable}{S\arabic{table}}
\renewcommand{\thesection}{S\arabic{section}}
\setcounter{figure}{0}
\setcounter{equation}{0}
\setcounter{section}{0}

\newcolumntype{C}[1]{>{\centering\arraybackslash}p{#1}}
\newcolumntype{L}[1]{>{\raggedright\arraybackslash}p{#1}}

\tableofcontents

\section*{Outline}
The supplementary information is divided into four sections.
In the first (Sec.~\ref{sec:hamiltonian_and_equiv_circuit}), we aim to derive the Hamiltonian of a voltage-biased mechanical oscillator coupled to a transmon. 
To do so, we will rely on an equivalent circuit of the mechanical part.
In the two following sections, we will derive the requirements to obtain phonon-number resolution in different parameter regimes.
We study both the coupling of motion to a single transmon (Sec.~\ref{sec:requirements_details_2body}), as well as to two electrical modes coupled through a cross-Kerr interaction (Sec.~\ref{sec:requirements_details_3body}).
Finally, we provide the numerical methods used to construct the simulations used to validate the theoretical derivations (Sec.~\ref{sec:numerical_methods}).

\section{Hamiltonian of a voltage-biased mechanical oscillator coupled to a transmon}
In this section, we will derive the Hamiltonian of a voltage-biased mechanical oscillator coupled to a transmon, displayed in the main text at Eq.~(\ref{eq:rabi_hamiltonian}).
We will first derive the equivalent circuit for the voltage-biased mechanical oscillator featured in Figs.~\ref{fig:drum_and_transmon} and~\ref{fig:equivalent_circuit}.
This will allow us to construct an circuit equivalent for the coupled system and thus make use of standard circuit quantization techniques to obtain the system Hamiltonian.

\label{sec:hamiltonian_and_equiv_circuit}
\subsection{Equivalent circuit of a voltage-biased mechanical oscillator}
\label{sec:equiv_circuit_drum}

\begin{figure}[b]
\centering
\includegraphics[width=0.75\textwidth]{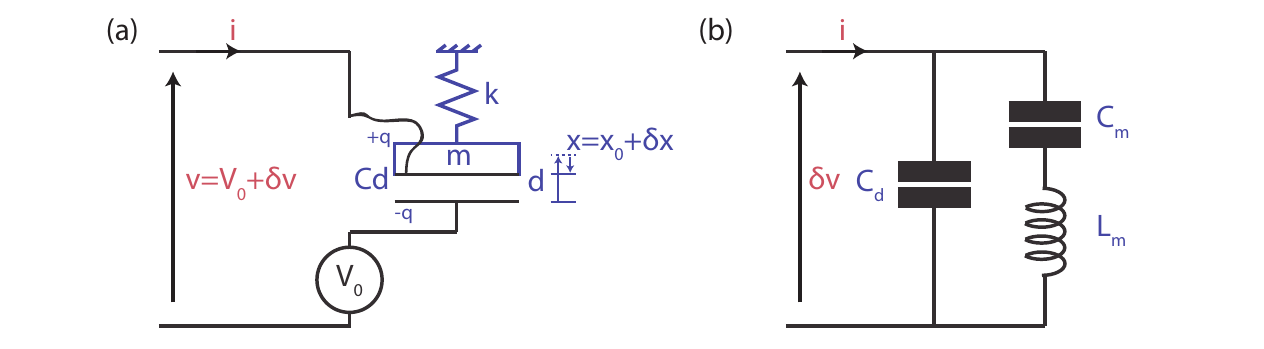}
\caption{
{\textbf{Equivalent circuit}. In panel (b) we show the equivalent circuit of the voltage-biased mechanical oscillator shown in panel (a).}
}
\label{fig:equivalent_circuit}
\end{figure}

Here we derive the equivalent circuit of a voltage-biased mechanical oscillator shown in Fig.~\ref{fig:equivalent_circuit}.
We will first derive the mechanical equations of motion, based on a Lagrangian description of the electromechanical system shown in Fig.~\ref{fig:equivalent_circuit}(a).
We will then obtain the electrical equations of motion, linking current injected towards the mechanical oscillator to voltage across it's capacitance.
After linearizing the equations around the static equilibrium imposed by the bias voltage, we will extract the admittance of the electro-mechanical system.
From this admittance, we can build an electrical circuit with the same admittance, which will constitute our equivalent circuit

\subsubsection{Mechanical equation of motion}
The Lagrangian of the system shown in Fig.~\ref{fig:equivalent_circuit}(a) is 
\begin{equation}
	\mathcal{L} = \frac{1}{2}C_d(x)v^2+\frac{1}{2}m\dot x^2 - \frac{1}{2}kx^2\ ,
	\label{eq:lagrangian}
\end{equation}
where the position dependent capacitance is
\begin{equation}
	C_d(x)=\frac{\epsilon_0A}{d-x}\ .
\end{equation}
The mechanical equation of motion is given by
\begin{equation}
	\frac{d}{dt}\left(\frac{\partial \mathcal{L}}{\partial\dot x}\right) = \frac{\partial \mathcal{L}}{\partial x}\ ,
	\label{eq:lagrange_eqs}
\end{equation}
yielding
\begin{equation}
	m\ddot x+kx=\frac{v^2}{2}\frac{\epsilon_0A}{(d-x)^2}\ .
	\label{eq:mech_eq_motion}
\end{equation}

\subsubsection{Electrical equation of motion}
From the definition of capacitance, we have
\begin{equation}
	q = C_d(x)v\ .
	\label{eq:qcv}
\end{equation}
By taking the derivative of this relation with respect to time, we get 
\begin{equation}
	\dot q = i = v\frac{dC_d(x)}{dt}+C_d(x)\dot v\ ,
\end{equation}
where
\begin{equation}
	\frac{dC_d(x(t))}{dt} = \epsilon_0A\frac{d}{dt}\left[\frac{1}{d-x(t)}\right] = \epsilon A \left[\frac{\dot x(t)}{(d-x(t))^2}\right]\ ,
\end{equation}
yielding the electrical equation of motion
\begin{equation}
	i = \dot v \frac{\epsilon_0A}{d-x}+v\dot x \frac{\epsilon_0A}{(d-x)^2}\ .
	\label{eq:elec_eq_motion}
\end{equation}

\subsubsection{Linearization around the static equilibrium}
We first write (\ref{eq:mech_eq_motion}) in a static limit $\frac{d}{dt}=0$ given an initial DC voltage $v = V_0$ leading to a constant displacement $x=x_0$
\begin{equation}
	kx_0=\frac{V_0^2}{2}\frac{\epsilon_0A}{(d-x_0)^2}\ .
	\label{eq:static}
\end{equation}
A DC voltage (positive or negative), will thus increase $x_0$, pulling the two plates of the capacitor closer together.
We rewrite (\ref{eq:mech_eq_motion}) considering small variations of voltage and position with respect to their static values $x\rightarrow x_0+\delta x$, $v\rightarrow V_0+\delta v$
\begin{equation}
	m\ddot {\delta x}+kx_0+k\delta x=\frac{(V_0+\delta v)^2}{2}\frac{\epsilon_0A}{(d-x_0-\delta x)^2}\ .
\end{equation}
By Taylor expanding to first order in $\delta x, \delta v$, and using Eq.~(\ref{eq:static}), we get
%
\begin{equation}
	m\ddot {\delta x}+k_\text{eff}(V_0)\delta x=\delta vV_0\frac{\epsilon_0A}{(d-x_0)^2}\ ,
	\label{eq:mech_eq_motion_pert}
\end{equation}
where the effective spring constant $k_\text{eff}(V_0)$ is
\begin{equation}
	k_\text{eff}(V_0) = k-\frac{V_0^2\epsilon_0A}{(d-x_0)^3}\ .
\end{equation}
By Taylor expanding the electrical equation of motion Eq.~(\ref{eq:elec_eq_motion}) to first order in $\delta x, \delta v$ around the static equilibrium, we obtain
\begin{equation}
	i = \dot {\delta v} \frac{\epsilon_0A}{d-x_0}+\dot {\delta x} V_0\frac{\epsilon_0A}{(d-x)^2}\ .
	\label{eq:elec_eq_motion_pert}
\end{equation}

\subsubsection{Frequency domain}

We introduce the phasors $\delta v\rightarrow  Ve^{j\omega t}$, $i\rightarrow  Ie^{j\omega t}$ and $\delta x\rightarrow   Xe^{j\omega t}$ where $ V,  X, I$ are time-independent complex number.
Additionally, we introduce the (voltage dependent) rest-position of the mechanical oscillator $D = d-x_0$ and the corresponding capacitance $C_d =\epsilon_0A/D$.
Making the above substitutions in Eqs~(\ref{eq:mech_eq_motion_pert}),(\ref{eq:elec_eq_motion_pert}), we get
\begin{equation}
	-\omega^2mX+k_\text{eff}X = V\frac{V_0C_d}{D}\ ,
	\label{eq:mech_eq_motion_sinu}
\end{equation}
\begin{equation}
	I = j\omega C_d V + j\omega C_d V_0\frac{X}{D}\ .
	\label{eq:elec_eq_motion_sinu}
\end{equation}
Eq.~(\ref{eq:mech_eq_motion_sinu}) gives the conversion between mechanical motion amplitude and the amplitude of voltage oscillations
\begin{equation}
	{X = V\frac{V_0C_d/D}{k_\text{eff}-\omega^2m}}\ .
\end{equation}
Plugged into Eq.~(\ref{eq:elec_eq_motion_sinu}), it provides the equivalent admittance of the voltage biased mechanical oscillator
\begin{equation}
	Y(\omega) = \frac{I}{V} = j\omega C_d + j\omega\frac{V_0^2C_d^2/D^2}{k_\text{eff}-\omega^2m}\ .
\end{equation}

\subsubsection{Equivalent circuit}
The admittance above can be rewritten 
\begin{equation}
	Y(\omega) = \frac{I}{V} = j\omega C_d + \frac{1}{\frac{1}{j\omega C_m}+j\omega L_m}\ ,
\end{equation}
which is exactly equal to the admittance of the circuit shown in Fig.~\ref{fig:equivalent_circuit}(b).
This circuit may thus serve as an equivalent circuit to the voltage biased mechanical oscillator.
The capacitance and inductance representing the mechanical mode are
\begin{equation}
	{C_m = \frac{V_0^2C_d^2}{D^2}\frac{1}{k_\text{eff}}}\ ,\ L_m = \frac{D^2}{V_0^2C_d^2}m\ .
\end{equation}

\subsection{Derivation of the system Hamiltonian}
\label{sec:hamiltonian_derivation}
In this section we will use circuit quantization~\cite{vool2017introduction} to obtain the Hamiltonian of the circuit of Fig.~\ref{fig:drum_and_transmon}(b). 
We define the flux $\phi$ from the voltage $v$ across a circuit component as
\begin{equation}
  \phi(t) = \int_{-\infty}^t v(t')dt'\ .
\end{equation}
By denoting the flux associated with the voltage across the inductor and the junction by $\phi_m$ and $\phi_t$ respectively, the Lagrangian of the system writes
\begin{equation}
\mathcal{L} = C_t\frac{\dot{\phi}_t^2}{2} + C_m\frac{(\dot{\phi}_t-\dot{\phi}_m)^2}{2}+E_J\cos\bigg(\frac{\phi_t}{\Phi_0}\bigg) - \frac{\phi_m^2}{2L_m}\ ,
\end{equation}
where $\Phi_0 = \hbar/2e$.
The canonical momenta (dimensionally charges) associated with the fluxes are given by
\begin{equation}
	q_t = \frac{\partial \mathcal{L} }{\partial \dot{\phi}_t} = (C_t+C_m)\dot\phi_t+C_m\dot\phi_m\ ,
\end{equation}
\begin{equation}
	q_m = \frac{\partial \mathcal{L} }{\partial \dot{\phi}_m} = C_m\dot\phi_t+C_m\dot\phi_m\ .
\end{equation}
Expressing the lagrangian using these variables rather than the derivative of flux yields
\begin{equation}
\mathcal{L} = \frac{1}{C_t}\frac{q_t^2}{2} + \frac{C_m+C_t}{C_mC_t}\frac{q_m^2}{2}+\frac{1}{C_t}q_mq_t+E_J\cos\bigg(\frac{\phi_t}{\Phi_0}\bigg) - \frac{\phi_m^2}{2L_m}\ .
\end{equation}
We now use a Legendre transformation to turn this Lagrangian into a Hamiltonian
\begin{equation}
\begin{split}
	H &= \sum_i q_i\dot{\phi}_i - \mathcal{L}\\
	&= \frac{1}{C_t}\frac{q_t^2}{2} + \frac{C_m+C_t}{C_mC_t}\frac{q_m^2}{2}+\frac{1}{C_t}q_mq_t-E_J\cos\bigg(\frac{\phi_t}{\Phi_0}\bigg) + \frac{\phi_m^2}{2L_m}\\
	&\simeq \frac{1}{C_t}\frac{q_t^2}{2} + \frac{C_m+C_t}{C_mC_t}\frac{q_m^2}{2}+\frac{1}{C_t}q_mq_t+\frac{\phi_t^2}{2L_J} - \frac{E_J}{24}\frac{\phi_t^4}{\Phi_0^4} + \frac{\phi_m^2}{2L_m}\  ,
\end{split}
\end{equation}
where the approximate equality holds in the limit of weak anharmonicity, or equivalently $\phi_t\ll \Phi_0$.
We have introduced the Josephson inductance as $L_J = E_J/\Phi_0^2$.
We now promote the classical variables to quantum variables $q_i\rightarrow\hat{q_i}$, $\phi_i\rightarrow\hat{\phi_i}$, postulating the commutation relation
\begin{equation}
  [\hat{\phi}_i,\hat{q}_j] = i\hbar\delta_{ij}\ .
\end{equation}
By introducing the associated creation and annihilation operators
\begin{equation}
\hat{\phi}_t = \sqrt{\frac{\hbar}{2}\sqrt{\frac{L_J}{C_t}}}(\hat{a}+\hat{a}^\dagger)\ ,\ \hat{q}_t = -i\sqrt{\frac{\hbar}{2}\sqrt{\frac{C_t}{L_J}}}(\hat{a}-\hat{a}^\dagger)\ ,
\end{equation}
\begin{equation}
\hat{\phi}_m = \sqrt{\frac{\hbar}{2}\sqrt{\frac{L_m(C_m+C_t)}{C_mC_t}}}(\hat{c}+\hat{c}^\dagger)\ ,\ \hat{q}_m = -i\sqrt{\frac{\hbar}{2}\sqrt{\frac{C_mC_t}{L_m(C_m+C_t)}}}(\hat{c}-\hat{c}^\dagger)\ ,
\end{equation}
we obtain the Hamiltonian of Eq.~(\ref{eq:rabi_hamiltonian}).

\section{Derivation of requirements for the coupling of a mechanical oscillator to a transmon}
\label{sec:requirements_details_2body}

In this section, we study the requirements for phonon-number resolution of a voltage biased mechanical oscillator coupled to a single transmon.
The Hamiltonian of interest is that of Eq.~(\ref{eq:rabi_hamiltonian}).
Our analysis is divided into three different regimes.
In Sec.~\ref{sec:max_chi} we study the case where the mechanical oscillator and transmon are very far detuned in frequency, such that the rotating-wave approximation (RWA) cannot be applied in analyzing the coupling $\Sigma\sim\Delta\sim\omega_t$.
In this regime we show that the magnitude of the cross-Kerr coupling $\chi_m$ has a maximum value depending on the mechanical frequency, which leads to the relation of Eq.~(\ref{eq:GHz_transmon_requirement}) discussed in the main text.
In Sec.~\ref{sec:RWA_regime} we study the case where the RWA may be applied, but where the mechanical oscillator and transmon are still detuned with respect to the coupling frequency, such that $\Sigma\gg\Delta\gg g$.
We show that this regime leads to more stringent requirements on the coupling magnitude than the resonant regime.
This resonant regime is studied in the final section (Sec.~\ref{eq:resonant_regime}), where we demonstrate that thermal strong coupling is a requirement for phonon-number resolution as discussed through Eq.~(\ref{eq:main_ineq_2body}) in the main text.

\subsection{Dispersive regime, non-RWA case \texorpdfstring{$\Sigma\sim\Delta\sim\omega_t$}{}}
\label{sec:max_chi}
Here we determine the condition to observe phonon-number resolution in the regime where the transmon and mechanical oscillator are very far detuned in frequency, such that $\Sigma\sim\Delta\sim\omega_t$, and the RWA which assumes $\Sigma\gg\Delta$ cannot be applied.
In this regime the systems is best described in the normal-mode basis of Eq.~(\ref{eq:bogo_hamiltonian}).
We will investigate the requirements to obtain phonon-number resolution both in the spectrum of the transmon (due to the cross-Kerr coupling $\chi_m$) and in the spectrum of the mechanical oscillator (due to its anharmonicity $\tilde A_m$).
\subsubsection{Requirement to attain \texorpdfstring{$\chi_m/\hbar\gg\gamma_t$}{}}
\label{sec:max_chi}
The quantity of interest is the cross-Kerr relative to the transmon frequency at this highest possible anharmonicity $A = \omega_t'/20$
\begin{equation}
	\chi_m/\hbar\omega_t' \simeq \frac{8}{20} g^2\frac{ \omega_m'^2}{\omega_t'^4}\ ,
\end{equation}
which should exceed the quality factor of the transmon $Q_t = \omega_t'/\gamma_t$.
Using Eq.~(\ref{eq:k_eff}) to express $\omega_m'$ and $g$ as a function of the un-biased and effective spring constants, we get
\begin{equation}
	\chi_m/\hbar\omega_t' \simeq \frac{1}{10}\left(\frac{\omega_m^0}{\omega_t'}\right)^3\sqrt{\frac{C_d^2(k-k_\text{eff})^2(C_d k+C_Jk_\text{eff})}{(C_J+C_d)^3k^3}}\ .
	\label{eq:chi_m_expanded}
\end{equation}
%
Using Mathematica, we find that the square-root
\begin{equation}
 	S = \sqrt{\frac{C_d^2(k-k_\text{eff})^2(C_d k+C_Jk_\text{eff})}{(C_J+C_d)^3k^3}}
 	\label{eq:S_def}
 \end{equation} 
may not exceed unity
\begin{equation}
\begin{split}
	\text{max}[S] &= \left(\frac{C_d}{C_J+C_d}\right)^{\frac{3}{2}}<1\\
	&\text{ at }k_\text{eff}=0 \text{ if }C_J<2C_d\ ,\\
	&= \frac{C_d}{15\sqrt{3}C_J}<1\\
	&\text{ at }k_\text{eff}=k\frac{C_J-2C_d}{3C_J} \text{ if }C_J>2C_d\ .
	\label{eq:max_S}
\end{split}
\end{equation}
By applying Eq.~(\ref{eq:max_S}) to Eq.~(\ref{eq:chi_m_expanded}), we find that
\begin{equation}
	\chi_m/\hbar\omega_t' < \frac{1}{10}\left(\frac{\omega_m^0}{\omega_t'}\right)^3\ ,
\end{equation}
and if we want phonon-number resolution $Q_t^{-1}\ll\chi_m/\hbar\omega_t'$, we obtain
\begin{equation}
	Q_t^{-1} \ll \frac{1}{10}\left(\frac{\omega_m^0}{\omega_t'}\right)^3\ ,
\end{equation}
which gives rise to the discussion of Eq.~(\ref{eq:GHz_transmon_requirement}) in the main text.
\subsubsection{Requirement to attain \texorpdfstring{$\tilde A_m/\hbar\gg\gamma_m^\text{eff}$}{}}
\label{sec:Am_greater_than_gammam}
Here, we will investigate the requirement to attain phonon-number resolution in the spectrum of the mechanical oscillator.
Due to its coupling to the transmon, the normal-mode corresponding to mostly mechanical oscillations acquires an anharmonicity $\tilde A_m$.
Following Ref.~\cite{gely2017nature}, the magnitude of this anharmonicity is related to the cross-Kerr interaction and the anharmonicity of the transmon through $\chi_m = 2\sqrt{\tilde A\tilde A_m}$, and can thus be approximated (assuming $\tilde A\sim A$) to
\begin{equation}
	\tilde A_m/\hbar = \frac{\chi_m^2}{4\tilde A}\simeq16(A/\hbar)g^4\frac{\omega_m'^4}{\omega_t'^8}\ .
\end{equation}
This quantity should exceed the linewidth of the mechanical mode, which will be broadened through it's interaction with the transmon to an effective linewidth $\gamma_m^\text{eff}$.
To determine this linewidth, we first consider that the dissipation of the transmons energy to the environment can be captured through a Lindblad operator~\cite{bishop2010circuit}
\begin{equation}
	\gamma_t\left(\hat a\rho\hat a^\dagger-\tfrac{1}{2}\hat a^\dagger\hat a\rho-\tfrac{1}{2}\rho\hat a^\dagger\hat a\right)
	\label{eq:transmon_lindblad_operator}
\end{equation}
where $\rho$ is the density matrix of the system.
The collapse operator $\hat a$ relates to the normal-mode collapse operators through~\cite{gely2017nature}
\begin{equation}
	\hat a\simeq \tilde a - 2 \frac{g}{\omega_t'}\tilde c\ .
\end{equation}
By injecting this expression into Eq.~(\ref{eq:transmon_lindblad_operator}), we note that the dissipation of the mode $\tilde c$ has a magnitude $4\gamma_t(g/\omega_t')^2$.
We will assume that this dissipation induced by the coupling dominates over the intrinsic dissipation rate of the mechanical oscillator, such that the effective dissipation rate of the mechanical mode is $\gamma_m^\text{eff} = 4\gamma_t(g/\omega_t')^2$.
Note that if phonon-number resolution is achieved in the mechanical spectrum, the effective linewidth will be even broader due to thermal effects (see Eq.~(S26) of Ref.~\cite{gely2019observation}).
But phonon-number resolution requires at least $\gamma_m^\text{eff}\ll\tilde A_m'/\hbar$, which writes
\begin{equation}
	4\gamma_t\frac{g^2}{\omega_t'^2}\ll16(A/\hbar)g^4\frac{\omega_m'^4}{\omega_t'^8}
\end{equation}
Considering the maximum anharmonicity which would maintain the transmon regime $A = \hbar\omega_t'/20$, this condition rewrites
\begin{equation}
	Q_t^{-1}\ll\frac{1}{20}\left(\frac{\omega_m'}{\omega_t'}\right)^5S
\end{equation}
where $S$ is defined in Eq.~(\ref{eq:S_def}) and has a maximum value of $1$ (see Eq.~(\ref{eq:max_S})).
We can rewrite the condition of phonon-number resolution as
\begin{equation}
	Q_t^{-1}\ll\frac{1}{20}\left(\frac{\omega_m'}{\omega_t'}\right)^5
\end{equation} 
which reveals that the requirements on the transmon dissipation rate is even more stringent than when striving for phonon-number resolution in the transmon spectrum.

\subsection{Dispersive regime, RWA case \texorpdfstring{$\Sigma\gg|\Delta|,|\Delta - A|\gg g$}{}}
\label{sec:RWA_regime}
Here, we determine the conditions to obtain phonon-number resolution in a regime closer to resonance.
Specifically, we now assume that the transmon and mechanical oscillator are close enough in frequency, that the RWA applies
\begin{equation}
	\Sigma\gg|\Delta|,|\Delta - A|\gg g\ .
\end{equation}
The cross-Kerr interaction is then given by~\cite{koch_charge-insensitive_2007}
\begin{equation}
	\chi_m = 2A\frac{g^2}{\Delta(\Delta-A/\hbar)}\ ,
\end{equation}
and the anharmonicity induced in the mechanics is
\begin{equation}
	A_m =  \chi_m^2/(4A) = A\frac{g^4}{\Delta^2(\Delta-A/\hbar)^2}
\end{equation}
following the relation $\chi_m = 2\sqrt{A_mA}$~\cite{gely2017nature}.
We now derive the conditions to obtain phonon-resolution either in the mechanical spectrum $A_m\gg\hbar \gamma_m^\text{eff}$ or in the transmon spectrum $\chi_m\gg \hbar \gamma_t^\text{eff}$.
\subsubsection{Requirements to attain \texorpdfstring{$A_m/\hbar\gg\gamma_m^\text{eff}$}{Am>>gamma m}}
The mechanical oscillator dissipation rate will be broadened through its interaction with the transmon, acquiring an effective dissipation rate
\begin{equation}
	\gamma_m^\text{eff} = \gamma_m + \gamma_t\frac{g^2}{\Delta(\Delta-A/\hbar)}\ ,
\end{equation}
which can be derived from Fermi's golden rule~\cite{koch_charge-insensitive_2007}.
The difference in powers of $(g/\Delta)$ between this dissipation rate and $A_m$ arise since the dissipation rate is proportional to the current traversing a resistor squared whilst the anharmonicity is to first order proportional to the fourth power of the current.
If mechanical Fock states are distinguishable in the spectrum of the mechanical oscillator, then the line-width will further be broadened by thermal effects.
We want the anharmonicity to be at least larger than the line-width of the second transition ($\ket{1}\leftrightarrow\ket{2}$), given by (see Eq.~(S26) of Ref.~\cite{gely2019observation})
\begin{equation}
	\left(3+8n_\text{th}\right)\left(\gamma_m + \gamma_t\frac{g^2}{\Delta(\Delta-A/\hbar)}\right)\ ,
\end{equation}
where $n_\text{th}$ is the average number of phonons in the mechanical oscillator due to its thermalization with the environment.
Assuming $n_\text{th}\gg1$, we then have requirements on the dissipation rate of both the mechanical oscillator and transmon
\begin{equation}
\begin{split}
&8n_\text{th}\gamma_m\ll A/\hbar\left(g^4/\Delta^2(\Delta-A/\hbar)^2\right)\ ,\\
&8n_\text{th}\gamma_t\ll A/\hbar\left(g^2/\Delta(\Delta-A/\hbar)\right)\ .
\end{split}
\end{equation}

\subsubsection{Requirements to attain \texorpdfstring{$\chi_m/\hbar\gg\gamma_t^\text{eff}$}{chi>>gamma t}}
Broadened by thermal effects, the effective transmon linewidth writes
\begin{equation}
	\gamma_t^\text{eff} = (1+4n_\text{th})\gamma_t
\end{equation}
(see Eq.~(S26) of Ref.~\cite{gely2019observation}).
The condition $\chi_m/\hbar \gg \gamma_t^\text{eff}$, assuming $n_\text{th}\gg 1$ roughly writes
\begin{equation}
2n_\text{th}\gamma_t\ll A/\hbar(g^2/\Delta(\Delta-A/\hbar))\ .
\end{equation}

\subsubsection{Conclusions}
Trying to reach $A_m\gg\gamma_m^\text{eff}$ or $\chi_m/\hbar\gg\gamma_t^\text{eff}$ leads to a similar requirement
\begin{equation}
(2,8)n_\text{th}\gamma_t\ll A/\hbar\left(g^2/\Delta(\Delta-A/\hbar)\right)
\end{equation}
with only varying constants (2,8).
However the dispersive regime conditions $g\ll |\Delta|$, $g\ll |\Delta - A|$ impose an upper bound on the term on the righ-hand side, such that $(A/\hbar)g^2/(\Delta(\Delta-A/\hbar))\ll g$.
To prove this, first impose $g < \epsilon |\Delta|$, $g < \epsilon|\Delta - A|$, where $\epsilon$ is a small quantity, which establishes a domain $D$. Then distinguish two cases: $2g/\epsilon <A$ and $2g/\epsilon >A$. In the former (latter) case $|\chi_m|$ has 4 (2) local maxima on the domain $D$ which are easy to find graphically. For each maximum, it is easy to prove that $|\chi_m|<4g\epsilon$.
We will thus summarize the requirements in the dispersive regime as follows
\begin{equation}
(2,8)n_\text{th}\gamma_t\lll g\ , 
\end{equation}
meaning that the requirements are even harsher than thermal strong coupling, which is required with resonant coupling (see following section).

\subsection{Resonant regime \texorpdfstring{$|\Delta|\ll g$}{delta << g}}
\label{eq:resonant_regime}
Here we derive the requirements to obtain phonon-number resolution of a resonantly coupled transmon and mechanical oscillator.
Through the analysis of two different parameter regimes $A\gg g$ or $g\gg A$, we find that thermal strong coupling is a requirement for phonon-resolution as written in Eq.~(\ref{eq:main_ineq_2body}) and discussed in the main text.
We will assume that the we can neglect the counter-rotating terms of the coupling
\begin{equation}
	-\hbar g (\hat a -\hat a^\dagger)(\hat c -\hat c^\dagger)\rightarrow -\hbar g (\hat a\hat c^\dagger -\hat a^\dagger\hat c)\ ,
\end{equation}
through the approximation $g\ll\omega_m',\omega_t'$

\subsubsection{Case \texorpdfstring{$A\gg \hbar g$}{A>>g}}
\begin{figure}
\centering
\includegraphics[width=0.75\textwidth]{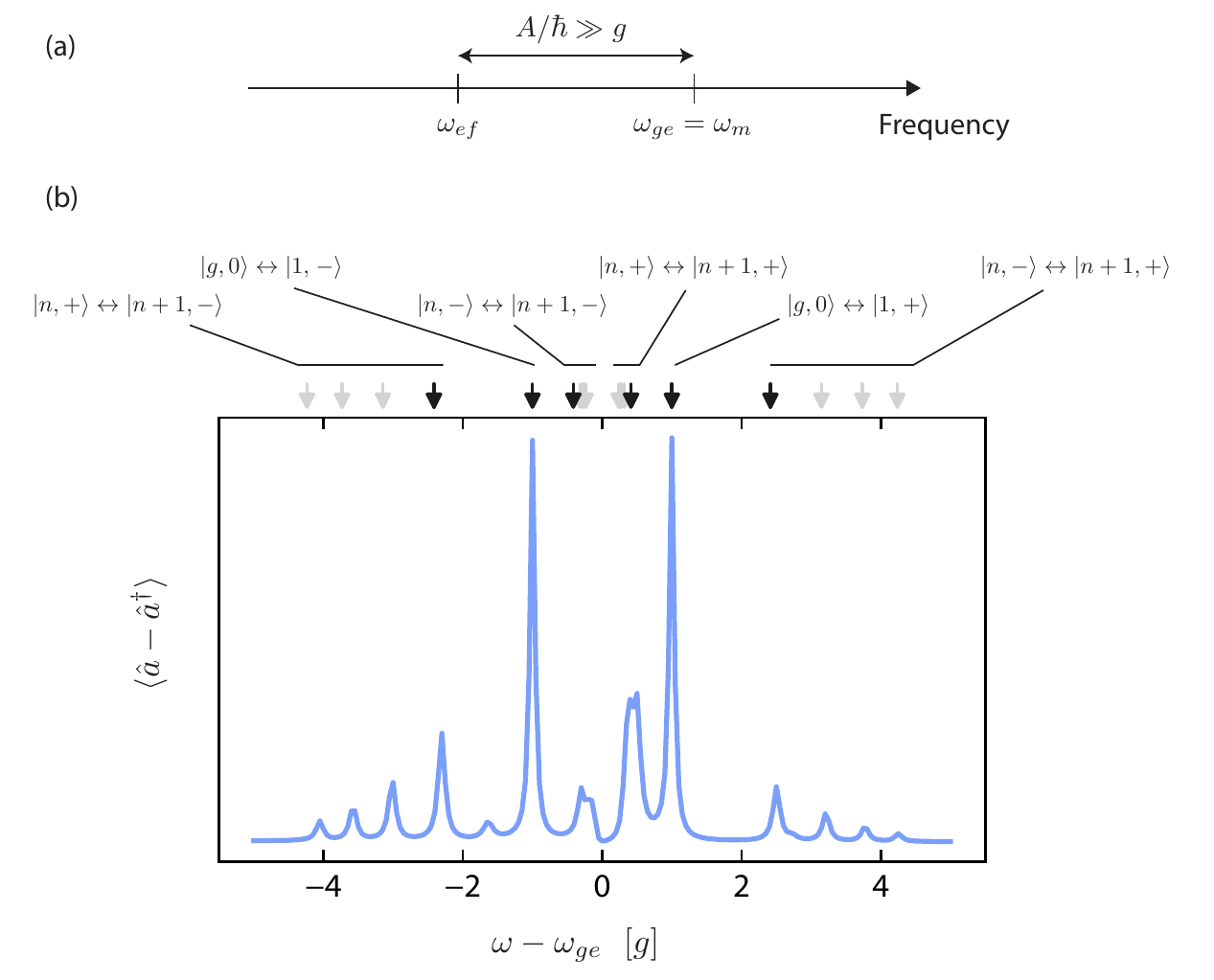}
\caption{
\textbf{Spectrum of an transmon coupled to a mechanical oscillator in the case }$A\gg \hbar g$.
{(a) Frequency landscape.}
We focus on the regime where the transmon anharmonicity $A$ dominates over the coupling $g$ such that the $\ket{e}\leftrightarrow\ket{f}$ transition of the transmon is effectively decoupled from the mechanical oscillator.
{(b) Spectrum.} 
We plot the expectation value $\langle\hat a - \hat a^\dagger\rangle$ whilst weakly driving the system at a frequency $\omega$.
This corresponds to one quadrature in a homodyne measurement of the transmon.
With the condition $2n_{\text{th}}\gamma_t\ll g/2$, we can resolve the Jaynes-Cummings spectrum of the coupled mechanical oscillator and transmon.
The exact simulation parameters and numerical methods are provided in Sec.~\ref{sec:numerical_methods}
}
\label{fig:g_A}
\end{figure}
In this case the transition frequency between the first and second excited state of the transmon is detuned from the mechanical oscillator frequency.
So we consider the mechanical oscillator to be coupled to a qubit constituted of the first two levels of the transmon $\ket{g}$ and $\ket{e}$.
In this regime, the eigenstates of the coupled system are that of the Jaynes-Cummings Hamiltonian~\cite{jaynes_comparison_1963}.
These are given by
\begin{equation}
	\ket{n,\pm} = \frac{\ket{g,n}\pm\ket{e,n-1}}{\sqrt{2}}\ ,
\end{equation}
with eigenenergies (where the groundstate $\ket{g,0}$ has 0 energy)
\begin{equation}
	\hbar\omega_{n,\pm} =  n\hbar\omega_{m,L}\pm\sqrt{n}\hbar g\ .
\end{equation}
Addressing the $\ket{0,g}\leftrightarrow \ket{1,\pm}$ transition independently of the $\ket{1,\pm}\leftrightarrow \ket{2,\pm}$ is our chosen definition of phonon-number resolution.
These transitions are separated in frequency by $g(2-\sqrt{2})\simeq g/2$.
The line-width of these transitions is the average of the line-width of the mechanical oscillator and transmon~\cite{Bishop2009a}, given by $(\gamma_m + \gamma_t(1+4 n_\text{th}))/2$ where the transmon line-width follows from Eq.~(S26) of Ref.~\cite{gely2019observation}.
Assuming the thermally broadened transmon line-width dominates over that of the mechanical oscillator and $n_\text{th}\gg1$, the condition for phonon-number resolution, such as in the spectrum of Fig.~\ref{fig:g_A}, is
\begin{equation}
	4\gamma_t n_\text{th} \ll g \ll A/\hbar\ .
	\label{eq:req_g_A}
\end{equation}

\subsubsection{Case \texorpdfstring{$\hbar g\gg A$}{g>>A}}
\label{sec:A_g}
\begin{figure}
\centering
\includegraphics[width=0.75\textwidth]{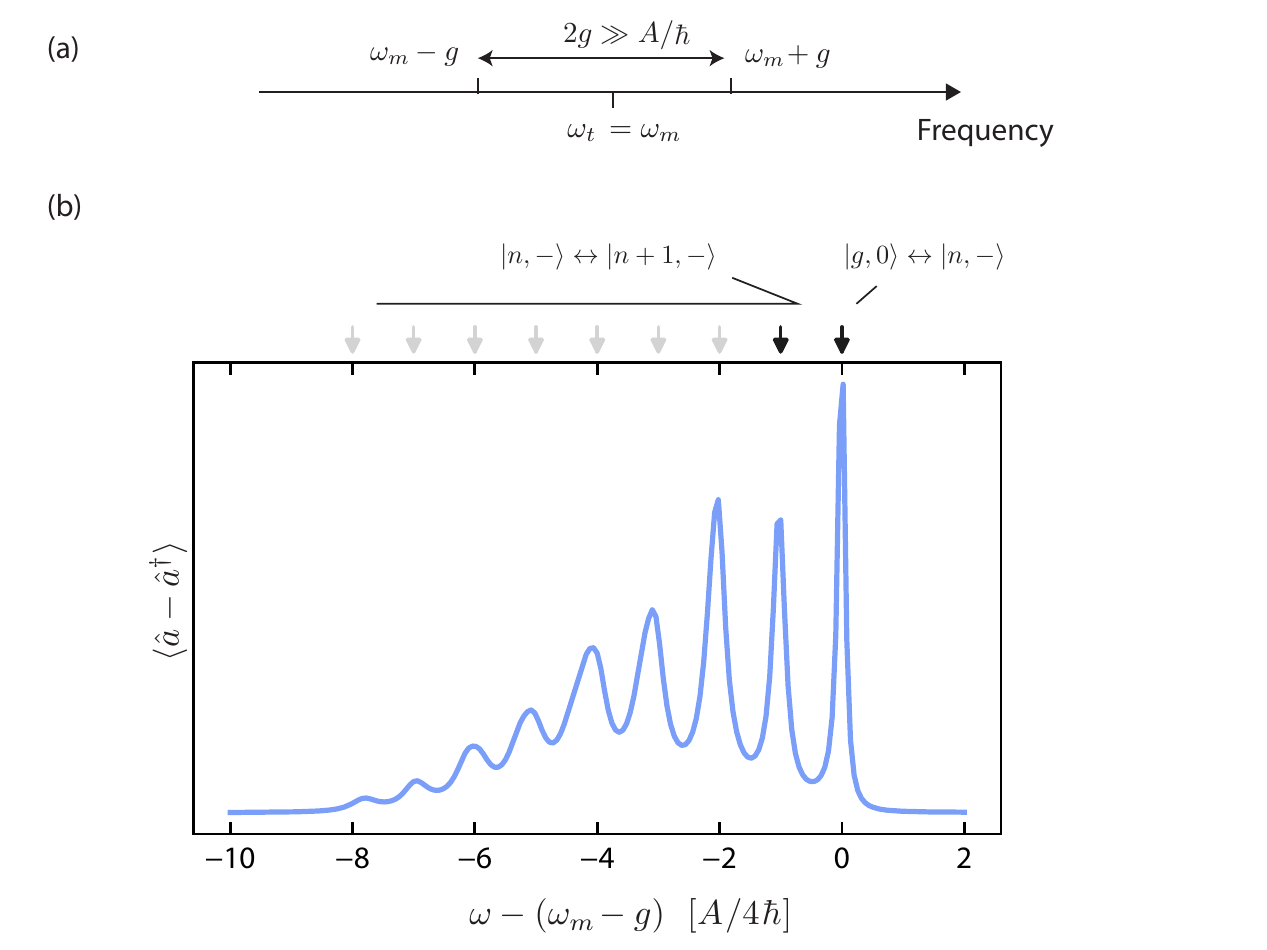}
\caption{
\textbf{Spectrum of an transmon coupled to a mechanical oscillator in the case} $A\ll \hbar g$.
{(a) Frequency landscape.} 
We focus on the regime where the mechanics-to-transmon coupling $g$ dominates over the transmon anharmonicity $A$ such that the two systems hybridize into two electromechanical modes separated in frequency by $2g$.
{(b) Spectrum.} 
We plot the expectation value $\langle\hat a - \hat a^\dagger\rangle$ whilst weakly driving the system at a frequency $\omega$.
This corresponds to one quadrature in a homodyne measurement of the transmon.
With the condition $4n_{\text{th}}\gamma_t/2\ll A/4\hbar$, we can resolve the anharmonic spectrum of one of the electro-mechanical modes.
The exact simulation parameters and numerical methods are provided in Sec.~\ref{sec:numerical_methods}
}
\label{fig:A_g}
\end{figure}

In this regime, the mechanical oscillator is close to resonance with multiple low-lying transitions of the transmon.
To capture this fact, we consider the anharmonicity of the transmon to be a perturbation on top of the harmonic terms of the transmon and the mechanical oscillator, which we will consider to be on resonance.
We call $\omega$ is the frequency of both the resonator and the harmonic part of the transmon Hamiltonian $\omega = \omega_m' = \omega_t'$.
We now perform a Bogoliubov transformation to the normal-mode basis, following Ref.~\cite{gely2017nature}.
Contrary to the calculations leading to the Hamiltonian of Eq.~(\ref{eq:bogo_hamiltonian}), resonance leads to normal-modes which are equally mechanical and electrical in nature, so we label their annihilation/creation operators and frequencies by $+$ and $-$.
Assuming $g\ll \omega$, the normal-modes are given by
\begin{align}
\begin{split}
\hat{a} &\simeq \frac{\hat{\alpha}_-+\hat{\alpha}_+}{\sqrt{2}} - \frac{g}{2\omega}\frac{\hat{\alpha}_-^\dagger-\hat{\alpha}_+^\dagger}{\sqrt{2}}\ ,\\
\hat{b} &\simeq \frac{\hat{\alpha}_--\hat{\alpha}_+}{\sqrt{2}} - \frac{g}{2\omega}\frac{\hat{\alpha}_-^\dagger+\hat{\alpha}_+^\dagger}{\sqrt{2}}\ ,\\
\tilde{\omega}_{+} &\simeq\omega+g-\frac{g^2}{2\omega}\ ,\\
\tilde{\omega}_{-} &\simeq\omega-g-\frac{g^2}{2\omega}\ ,\\
\label{eq:chapter-2_normal_mode_resonant}
\end{split}
\end{align}
to first order in $g/\omega$.
The Hamiltonian in the normal-mode basis is then
\begin{equation}
  \hat H = \hbar\omega_+\hat{\alpha}_+^\dagger\hat{\alpha}_++\hbar\omega_-\hat{\alpha}_-^\dagger\hat{\alpha}_--\frac{E_c/4}{12}(\hat \alpha_+ +\hat \alpha_+^\dagger+\hat \alpha_- +\hat \alpha_-^\dagger)^4\ .
\end{equation}
Expanding the quartic term, and keeping only terms which will be relevant in first order perturbation theory, leads to
\begin{align}
\begin{split}
  \hat{H}\simeq&\hbar\omega_+\hat{\alpha}_+^\dagger\hat{\alpha}_++\hbar\omega_-\hat{\alpha}_-^\dagger\hat{\alpha}_-\\
  &-\frac{ A_{+}}{2}\left(\left(\hat{\alpha}_+^\dagger\hat{\alpha}_+\right)^2 +\hat{\alpha}_+^\dagger\hat{\alpha}_++ \frac{1}{2}\right)\\
  &-\frac{ A_{-}}{2}\left(\left(\hat{\alpha}_-^\dagger\hat{\alpha}_-\right)^2 +\hat{\alpha}_-^\dagger\hat{\alpha}_-+ \frac{1}{2}\right)\\
  &-\chi\left(\hat{\alpha}_+^\dagger\hat{\alpha}_++\frac{1}{2}\right)\left(\hat{\alpha}_-^\dagger\hat{\alpha}_-+\frac{1}{2}\right)\ ,\\
  \label{eq:chapter-2_resonant_Hamiltonian_large_g}
\end{split}
\end{align}
with anharmonicities $A_\pm = E_C/4$ and a cross-Kerr interaction $\chi = E_C/2$.
%
%
%
%
Following Fermi's golden rule, the modes will have dissipation rates $(\gamma_t+\gamma_m)/2$~\cite{koch_charge-insensitive_2007}.
To resolve at least the transition between the ground and first-excited state of the electro-mechanical modes, the line-width of the first-to-second excited state, dressed by thermal effects should be smaller than the mode anharmonicity.
Assuming the transmon has a dominating dissipation rate, and that the modes are thermally populated $n_\text{th}\gg1$, the larger line-width of the first-to-second excited state is given by (see Eq.~(S26) of Ref.~\cite{gely2019observation})
\begin{equation}
	\gamma_\pm\simeq 8n_\text{th}\gamma_t .
\end{equation}
Phonon-number resolution is then achievable when
\begin{equation}
8n_\text{th}\gamma_t\ll A/\hbar \ll g\ .
\end{equation}
An example of the obtainable spectrum is given in Fig.~\ref{fig:A_g}.
This requirement is combined with that of Eq.~(\ref{eq:req_g_A}) in the requirement quoted in the main text in Eq.~(\ref{eq:main_ineq_2body}).

\section{Derivation of requirements for coupling a mechanical oscillator to two superconducting modes}{}
\label{sec:requirements_details_3body}
We now consider adding a second high-frequency (HF) mode.
In this situation, we analyse different regime of parameters, to derive the requirement for phonon-number resolution of Eq.~(\ref{eq:main_ineq_3body}) discussed in the main text.
The HF mode is assumed to to be coupled (through the junction non-linearity) by a cross-Kerr interaction to the low-frequency (LF) electrical mode, in the spirit of Ref.~\cite{gely2019observation}.
The LF mode is still assumed to be resonant with the mechanical oscillator.
The system follows the Hamiltonian of Eq.~(\ref{eq:two-mode-hamiltonian}).
We have neglected many of the terms which arise from the quartic non-linearity under the assumptions $A_H\ll\hbar \omega_{t,H}'$ and $A_L,\chi_{LH}\ll\hbar \omega_{L,m}$.
Note that we necessarily have $\chi_{LH} = 2\sqrt{A_HA_L}$.
We will additionally assume that the we can neglect the counter-rotating terms of the coupling
\begin{equation}
	-\hbar g (\hat b -\hat b^\dagger)(\hat c -\hat c^\dagger)\rightarrow -\hbar g (\hat b\hat c^\dagger -\hat b^\dagger\hat c)\ ,
\end{equation}
through the approximation $g\ll\omega_m',\omega_{t,L}'$

\subsection{Case \texorpdfstring{$A_L\gg\hbar  g\gg \chi_{LH}$}{Al>>g>>chi}}
\begin{figure}
\centering
\includegraphics[width=0.75\textwidth]{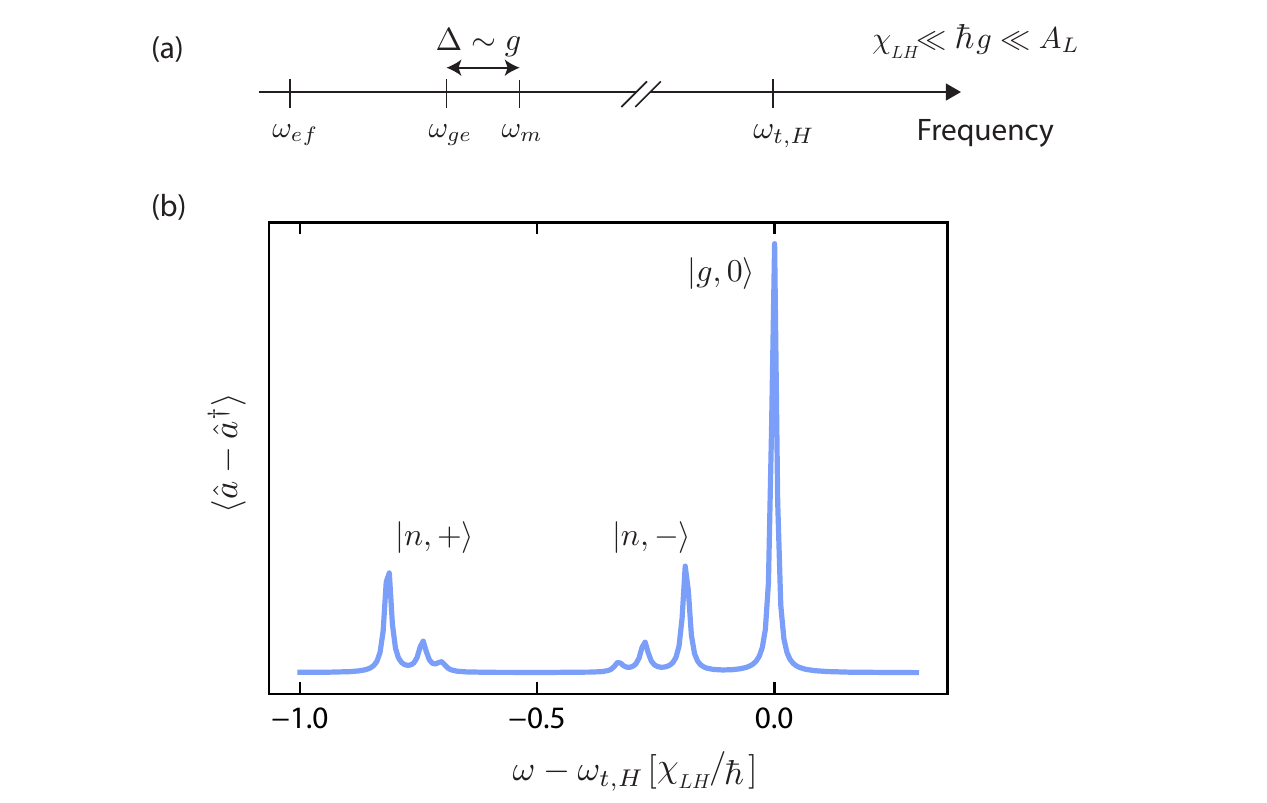}
\caption{
\textbf{HF mode spectrum in the case} $A_L\gg \hbar g\gg \chi_{LH}$.
{(a) Frequency landscape.}
In the regime where $A_L$ dominates over the coupling rate, the $\ket{e}\leftrightarrow\ket{f}$ transition of the LF mode is effectively decoupled of the mechanical oscillator.
The near resonant oscillator and $\ket{g}\leftrightarrow\ket{e}$ transition of the LF mode then hybridize following Jaynes-Cummings physics.
{(b) Spectrum.} 
We plot the expectation value $\langle\hat a - \hat a^\dagger\rangle$ whilst weakly driving the system at a frequency $\omega$.
This corresponds to one quadrature in a homodyne measurement of the HF mode.
With the condition $\gamma_H + 4n_{\text{th}}\gamma_t\ll \chi_{LH}/\hbar/15$, we can resolve phonon-number dependent transitions.
The exact simulation parameters and numerical methods are provided in Sec.~\ref{sec:numerical_methods}
}
\label{fig:chi_g_A}
\end{figure}
In the assumption $A_L\gg\hbar  g$, the first-to-second-excited state transition of the LF  mode will be detuned and uncoupled to the mechanical oscillator.
The mechanical oscillator is then effectively coupled to a qubit constituted of the ground $\ket{g}$ and excited $\ket{e}$ states of the LF mode.
We may thus rewrite the Hamiltonian as
\begin{equation}
\begin{split}
	\hat H &= \hbar \omega_{t,H}' \hat a ^\dagger \hat a - A_H\hat a ^\dagger\hat a ^\dagger \hat a \hat a \\
	&+\sum_{j\ge 2}(E_j-j\chi_{LH} \hat a ^\dagger\hat a) \ket{j}\bra{j}\\
	&+ \hbar \omega_{t,L}' \ket{e}\bra{e} + \hbar (\omega_{t,L}'+\Delta)\hat c ^\dagger \hat c \\
	&+ \hbar g (\ket{g}\bra{e}\hat c ^\dagger +\ket{e}\bra{g}\hat c )- \sum_n\chi_{LH} \hat a ^\dagger \hat a \ket{e,n}\bra{e,n}\ .
\end{split}
\end{equation}
Note that multiplied the last term by identity $\sum_n\ket{n}\bra{n}$.
We may move to the eigen-basis of the Jaynes-Cummings Hamiltonian which will capture the coupling between the LF mode qubit and the mechanical oscillator~\cite{jaynes_comparison_1963,bishop2010circuit}
\begin{equation}
\begin{split}
\ket{n,+}&= \cos \theta_n\ket{e,n-1} + \sin \theta_n\ket{g,n}\ ,\\
\ket{n,-}&= -\sin \theta_n\ket{e,n-1} + \cos \theta_n\ket{g,n}\ ,\\
\tan 2\theta_n &= -2g\sqrt{n}/\Delta\ ,\\
\end{split}
\end{equation}
where the states $\ket{n,\pm}$ have energies
\begin{equation}
\omega_{n,\pm} =  (\omega_{t,L}'+\Delta) n-\frac{\Delta}{2} \pm  \frac{1}{2}\sqrt{4g^2 n+\Delta^2}\ .
\end{equation}  
Constant energy terms are subtracted such that the ground state $\ket{0,g}$ has 0 energy.
The resulting Hamiltonian is
\begin{equation}
\begin{split}
	\hat H &= \hbar \omega_{t,H}' \hat a ^\dagger \hat a - A_H\hat a ^\dagger\hat a ^\dagger \hat a \hat a  \\
	&+\sum_{j\ge 2}(E_j-j\chi_{LH} \hat a ^\dagger\hat a) \ket{j}\bra{j}\\
	&+ \sum_{n\ge1,s=\pm}\hbar \omega_{n,s}\ket{n,s}\bra{n,s} \\
	&-\chi_{LH}\sum_{n\ge1} \cos(\theta_n)^2\hat a ^\dagger \hat a\ket{n,+}\bra{n,+}\\
	&-\chi_{LH}\sum_{n\ge1} \sin(\theta_n)^2\hat a ^\dagger \hat a\ket{n,-}\bra{n,-}\ ,
\end{split}
\end{equation}
where we neglected the term
\begin{equation}
\chi_{LH}\cos(\theta_n)\sin(\theta_n)\sum_ n \left(\ket{n,+}\bra{n,-}+\ket{n,-}\bra{n,+}\right)\ ,
\end{equation}
valid in the limit which couple terms separated in frequency by $\sqrt{4g^2 n+\Delta^2}$ which is much smaller than $\chi_{LH}$ under the initial assumption $\chi_{LH}\ll\hbar g$.
The HF spectrum reveals the transitions
\begin{equation}
	\ket{g,n,\pm}\leftrightarrow \ket{e,n,\pm}\ ,
\end{equation}
at frequencies
\begin{equation}
	\omega_{t,H}'-\chi_{n,\pm}/\hbar\ ,
\end{equation}
where $\chi_{n,+}  = \chi_{LH}\cos(\theta_n)^2$ and $\chi_{n,-}  = \chi_{LH}\sin(\theta_n)^2$.
Different values of the ratio $g/\Delta$ lead to different frequencies, note that we always have $0<\chi_{n,\pm}<\chi_{LH}$.
If $g/\Delta\ll 1$, $\chi_{n,-}\sim0$ and $\chi_{n,+}\sim \chi_{LH}$, and the HF mode is insensitive to $n$ and only sensitive to the states $\ket{g},\ket{e}$ of the LF mode.
If $g/\Delta\gg 1$, $\chi_{n,\pm}\sim\chi_{LH}/2$, and again sensitivity to $n$ is lost.
%
Numerically, we find that both $|\chi_{1,+}-\chi_{2,+}|$ and  $|\chi_{1,-}-\chi_{2,-}|$ have a maximum at $\sim \chi/15$ for $g/\Delta\simeq 0.6$ yielding
\begin{equation}
	\begin{split}
\chi_{1,+}&= 0.82\chi_{LH}\\
\chi_{1,-}&= 0.18\chi_{LH}\\
\chi_{2,+}&= 0.75\chi_{LH}\\
\chi_{2,-}&= 0.25\chi_{LH}\ .\\
	\end{split}
\end{equation}
In order to resolve the $\ket{g,1,\pm}\leftrightarrow \ket{e,1,\pm}$ transitions, the detuning to the transition $\ket{g,2,\pm}\leftrightarrow \ket{e,2,\pm}$, should exceed the line-width of the latter transition that we will call $\gamma_{H,\text{eff}}$
\begin{equation}
	\gamma_{H,\text{eff}}\ll\chi_{LH}/15\hbar\ .
\end{equation}
An example of the obtainable spectrum is given in Fig.\ref{fig:chi_g_A}.

\subsection{Case \texorpdfstring{$A_L,\chi_{LH}\gg \hbar g $}{Al,chi>>g}}
\begin{figure}
\centering
\includegraphics[width=0.75\textwidth]{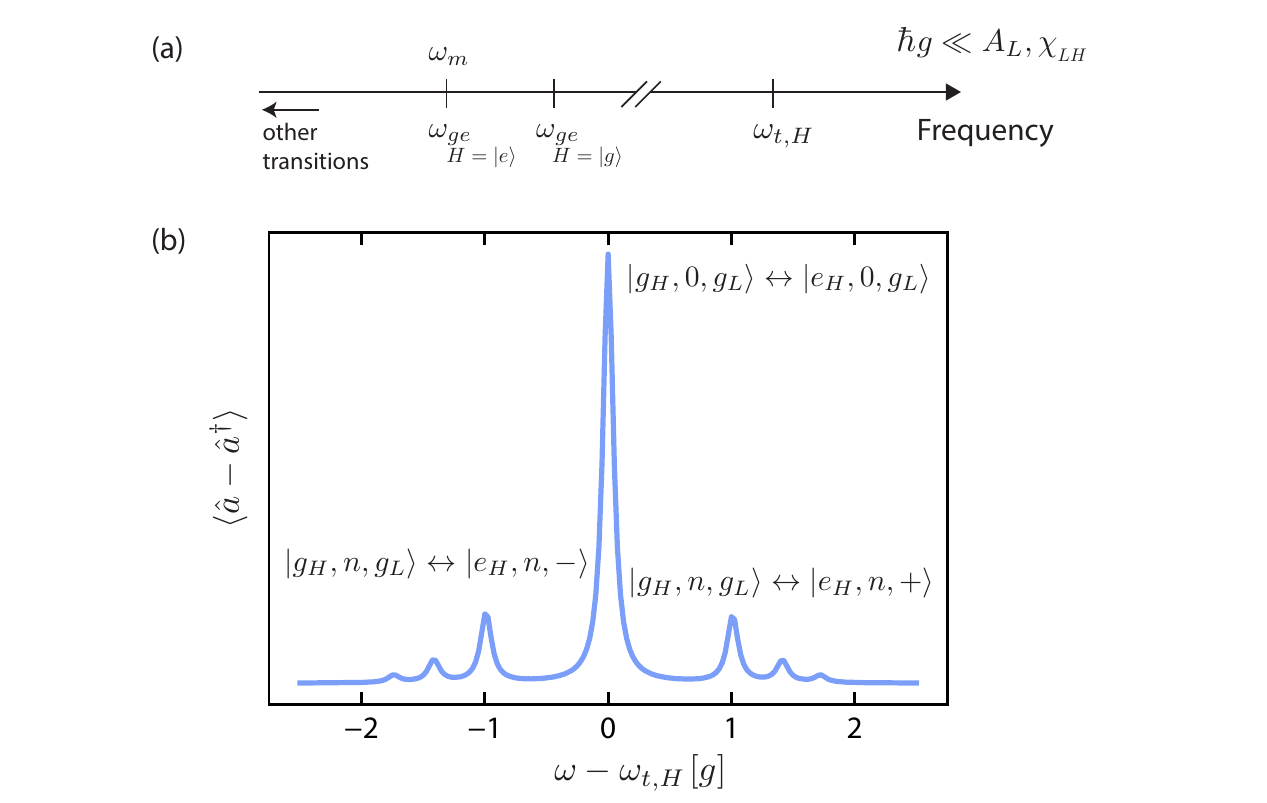}
\caption{
\textbf{HF mode spectrum in the case $A_L,\chi_{LH}\gg \hbar g $. }
{(a) Frequency landscape.}
In the regime where $\chi_{LH}$ dominates the mechanics-to-low-mode coupling $g$, the LF mode can only be coupled to the mechanical oscillator for a given state of the HF mode.
As when the HF mode is in a different state, the frequency of the LF mode is shifted by $\chi_{LH}\gg \hbar g$.
We focus on the case where the mechanical oscillator is resonant with the LF mode if the HF mode is in state $\ket{e}$.
Additionally, if $A_L$ dominates over the coupling rate, the $\ket{e}\leftrightarrow\ket{f}$ transition of the LF mode is effectively decoupled of the mechanical oscillator.
The near resonant oscillator and $\ket{g}\leftrightarrow\ket{e}$ transition of the LF mode then hybridize when the HF mode is in its excited state following Jaynes-Cummings physics.
{(b) Spectrum.} 
We plot the expectation value $\langle\hat a - \hat a^\dagger\rangle$ whilst weakly driving the system at a frequency $\omega$.
This corresponds to one quadrature in a homodyne measurement of the HF mode.
With the condition $\gamma_{t,H} + 4n_{\text{th}}\gamma_{t,L}\ll g/2$, we can resolve phonon-number dependent transitions at frequencies $\omega_{t,H}' \pm g\sqrt{n}$, where $n$ is the number of phonons in the mechanical oscillator.
The exact simulation parameters and numerical methods are provided in Sec.~\ref{sec:numerical_methods}
}
\label{fig:g_A_chi}
\end{figure}
We write the Hamiltonian in the basis of eigenstates of the HF and LF modes.
\begin{equation}
\begin{split}
	\hat H =& \sum_{j_H}\ket{j_H}\bra{j_H}\otimes\left[E_{j_H}\right.\\
	+& \sum_{j_L}(E_{j_L}-j_H\chi_{LH}) \ket{j_L}\bra{j_L} +  (\hbar\omega_{t,L}'-\chi_{LH}) \hat c ^\dagger \hat c\\
	-& g(\ket{j_L+1}\bra{j_L} \hat c+\ket{j_L}\bra{j_L+1}  \hat c^\dagger)\left.\right]\ .
\end{split}
\end{equation}
Here the frequency of the LF mode depends on the state of the HF mode.
Since $\chi_{LH}\gg \hbar g$, the LF mode will only be resonant with the mechanical oscillator when the HF mode is in a specific state.
We study the case where the LF mode and mechanical mode are resonant if the HF frequency mode is in its excited state.
Additionally, since $A_L\gg \hbar g$, only a single transition of the LF mode will be on resonance with the mechanical oscillator, the ground to excited state transition in this case.
We can thus rewrite the Hamiltonian as
\begin{equation}
\begin{split}
	\hat H = \ket{g_H}\bra{g_H}\otimes&\bigg[\sum_{j_L} E_{j_L} \ket{j_L}\bra{j_L} +  (\hbar\omega_{t,L}'-\chi_{LH}) \hat c ^\dagger \hat c \bigg]\\
	+\ket{e_H}\bra{e_H}\otimes&\bigg[\hbar\omega_{t,H}' + \sum_{j_L\ge2} (E_{j_L}-\chi_{LH}) \ket{j_L}\bra{j_L} \\
	&+(\hbar\omega_{t,L}'-\chi_{LH}) \ket{e_L}\bra{e_L}+  (\hbar\omega_{t,L}'-\chi_{LH}) \hat c ^\dagger \hat c \\
	&- g(\ket{e_L}\bra{g_L} \hat c+\ket{g_L}\bra{e_L}  \hat c^\dagger)\bigg]+...\\
\end{split}
\end{equation}
We now apply the unitary transformation $\Sigma_{j_H}\hat U_{j_H}$, where $U_{j_H}$ is identity except for $j_H=1$, when it brings the LF mode and mechanical oscillator to the Jaynes-Cummings basis
\begin{equation}
\ket{n,\pm}= \left(\ket{g_L,n}\pm \ket{e_L,n-1}\right)/\sqrt{2}\ ,\\
\end{equation}
where the states $\ket{n,\pm}$ have energies $\hbar\omega_{n,\pm}$
\begin{equation}
\omega_{n,\pm} =  (\omega_{t,L}'-\chi_{LH}/\hbar) n \pm g\sqrt{n}\ .
\end{equation}  
Constant energy terms are subtracted such that the ground state $\ket{0,g}$ has 0 energy.
The Hamiltonian writes
\begin{equation}
\begin{split}
	\hat H = \ket{g_H}\bra{g_H}\otimes&\bigg[\sum_{j_L} E_{j_L} \ket{j_L}\bra{j_L} + \hbar (\omega_{t,L}'+\Delta) \hat c ^\dagger \hat c \bigg]\\
	+\ket{e_H}\bra{e_H}\otimes&\bigg[\hbar\omega_{t,H}' + \sum_{j_L\ge2} E_{j_L} \ket{j_L}\bra{j_L} \\
	&+\sum_{n,s=\pm}\hbar\omega_{t,L}'\ket{n,s}\bra{n,s} \bigg] + ...\\
\end{split}
\end{equation}
Probing the HF spectrum will reveal the following transitions
\begin{equation}
\begin{split}
	\ket{g_H,n,g_L}\leftrightarrow \ket{e_H,n,\pm}\ ,\\
	\ket{g_H,n,e_L}\leftrightarrow \ket{e_H,n+1,\pm}\ ,
\end{split}
\end{equation}
with frequencies
\begin{equation}
\begin{split}
&(\omega_{t,H}' + \omega_{n,\pm})-(n(\omega_{t,L}'-\chi_{LH}/\hbar))\ ,\\
&(\omega_{t,H}' + \omega_{n+1,\pm})-(n(\omega_{t,L}'-\chi_{LH}/\hbar)+\omega_{t,L}')\\
\end{split}
\end{equation}
\begin{equation}
\begin{split}
&\omega_{t,H}' \pm g\sqrt{n}\ ,\\
&\omega_{t,H}' -\chi_{LH}/\hbar \pm g\sqrt{n}\\
\end{split}
\end{equation}
In order to resolve the $\ket{g,1,\pm}\leftrightarrow \ket{e,1,\pm}$ transitions, the detuning to the transition $\ket{g,2,\pm}\leftrightarrow \ket{e,2,\pm}$, given by $g\sqrt{2}-g\sim g/2$, should exceed the line-width of the latter transition $\gamma_{H,\text{eff}}$
\begin{equation}
	\gamma_{H,\text{eff}}\ll g/2\ .
\end{equation}
An example of the obtainable spectrum is given in Fig.\ref{fig:g_A_chi}.

\subsection{Case \texorpdfstring{$\hbar g\gg \chi_{LH}\gg A_L$}{g>>chi>>Al}}

In this regime, we write the LF mode as harmonic
\begin{equation}
\begin{split}
	\hat H &= \hbar \omega_{t,H}' \hat a ^\dagger \hat a - A_H\hat a ^\dagger\hat a ^\dagger \hat a \hat a \\
	&+ \hbar \omega_{t,L}' \hat b ^\dagger \hat b+ \hbar (\omega_{t,L}'+\Delta) \hat c ^\dagger \hat c\\
	&+ \hbar g (\hat b\hat c ^\dagger +\hat b^\dagger\hat c )- \chi_{LH} \hat a ^\dagger\hat a \hat b ^\dagger \hat b\ ,
\end{split}
\end{equation}
which comes under the condition that its anharmonicity represents a perturbation to the Hamiltonian smaller than the other interaction rates.
\begin{figure}
\centering
\includegraphics[width=0.75\textwidth]{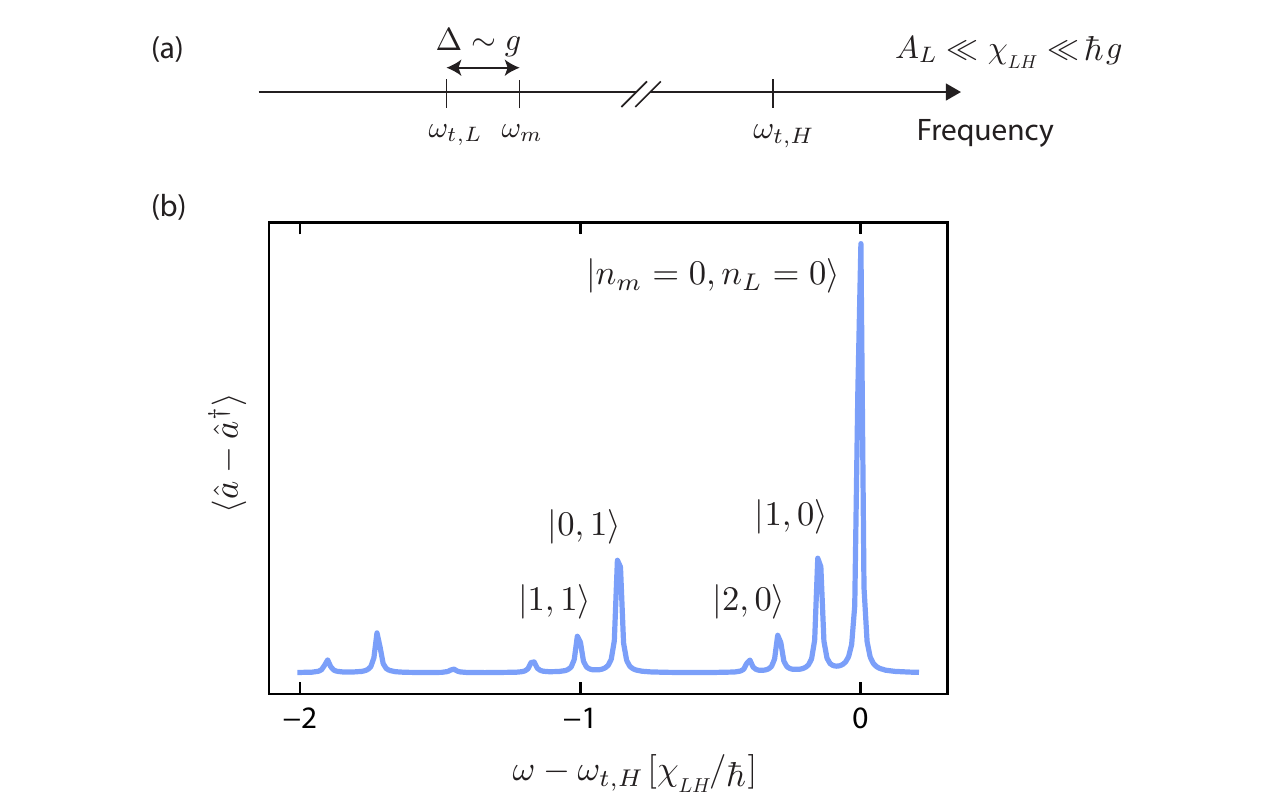}
\caption{
\textbf{HF mode spectrum in the case} $\hbar g\gg \chi_{LH}\gg A_L$ .
{(a) Frequency landscape.}
In the regime where the mechanics-to-low-mode coupling $\hbar g$ dominates over the LF mode anharmonicity $A_L$, the LF mode and mechanical oscillator hybridize into two electromechanical modes.
{(b) Spectrum.} 
We plot the expectation value $\langle\hat a - \hat a^\dagger\rangle$ whilst weakly driving the system at a frequency $\omega$.
This corresponds to one quadrature in a homodyne measurement of the HF mode.
With the condition $\gamma_{t,H} + 4n_{\text{th}}\gamma_{t,L}\ll \chi_m/\hbar$, we can resolve phonon-number dependent transitions at frequencies $\omega_{t,H}' - n_L\chi_L/\hbar- n_m\chi_m/\hbar$, where $n_m$ is the number of phonons in the most mechanical electromechanical mode, and $n_L$ the number of photons in the most electrical electro-mechanical mode, and $\chi_m+\chi_L=\chi_{LH}$.
The exact simulation parameters and numerical methods are provided in Sec.~\ref{sec:numerical_methods}
}
\label{fig:A_chi_g}
\end{figure}
We first perform a first basis change to the normal-modes resulting from the $g$ coupling, leading to electromechanical modes indexed by $1$, with annihilation operators defined by
\begin{equation}
\begin{split}
\hat b &= f(g/\Delta)\hat b_1 + h(g/\Delta) \hat c_1\ ,\\
\hat c &= f(g/\Delta)\hat c_1 - h(g/\Delta) \hat b_1\ ,
\label{eq:mode_1_annihilation}
\end{split}
\end{equation}
where
\begin{equation}
\begin{split}
	f(x) &=  \frac{1+\sqrt{1+4x^2}}{\sqrt{8x^2+2\left(1+\sqrt{1+4x^2}\right)}}\ , \\
	h(x) &= \frac{2x}{\sqrt{8x^2+2\left(1+\sqrt{1+4x^2}\right)}}\ ,
\end{split}
\end{equation}
and mode frequencies
\begin{equation}
\begin{split}
\omega_1+\Delta_1 &= \omega_{t,L}'+\frac{\Delta}{2}\left(1+\sqrt{1+4\frac{g^2}{\Delta^2}}\right)\ ,\\
\omega_1 &= \omega_{t,L}'+\frac{\Delta}{2}\left(1-\sqrt{1+4\frac{g^2}{\Delta^2}}\right)\ ,\\
\label{eq:mode_1_freqs}
\end{split}
\end{equation}
as calculated with the Bogoliubov transformation described in Ref.~\cite{gely2017nature}.
The resulting Hamiltonian is
\begin{equation}
\begin{split}
	\hat H &= \hbar \omega_{t,H}' \hat a ^\dagger \hat a - A_H\hat a ^\dagger\hat a ^\dagger \hat a \hat a \\
	&+ \hbar \omega_1 \hat b_1^\dagger \hat b_1+ \hbar (\omega_1+\Delta_1) \hat c_1 ^\dagger \hat c_1\\
	&- \chi_{LH} f(g/\Delta)^2 \hat a ^\dagger\hat a \hat b_1 ^\dagger \hat b_1 - \chi_{LH} h(g/\Delta)^2 \hat a ^\dagger\hat a \hat c_1 ^\dagger \hat c_1\ ,
\end{split}
\label{eq:mode_1_Hamiltonian_full}
\end{equation}
where we neglected the term
\begin{equation}
	- \chi_{LH} f(g/\Delta)h(g/\Delta)\hat a ^\dagger\hat a(\hat b_1 ^\dagger \hat c_1+\hat b_1  \hat c_1^\dagger)\ ,
\end{equation}
valid in the limit $\hbar g\gg\chi_{LH}$ where the interaction $\hat b_1 ^\dagger \hat c_1$ couples states separated in energy by $2g$ (near resonance), assumed to be much larger than the strength of this interaction, $\chi_{LH} f(g/\Delta)h(g/\Delta)<\chi_{LH}$.
We now look for a reasonable choice for the parameter $g/\Delta$.
In the case $\Delta=0$, the cross-Kerr interaction between the electromechanical modes $\chi_{LH} f(g/\Delta)^2$ and $\chi_{LH} h(g/\Delta)^2$ will be identical and equal to $\chi_{LH}/2$.
The HF spectrum will then feature the transition frequencies
\begin{equation}
	\hbar\omega_{t,H}'-\chi_{LH}(n_++n_-)/2\ ,
\end{equation}
where $n_\pm$ refers to the occupation of each electromechanical mode.
Each measured peak would then correspond to multiple states, which restricts the level of control attainable over the quantum states of each electromechanical mode.
Alternatively, with $g/\Delta\ll1$, the LF mode would weakly hybridize, with a very small cross-Kerr coupling to the more mechanical degree of freedom.
An in-between is thus desireable, with $g/\Delta \sim 1$ giving rise to two electromechanical modes, one dominantly electrical, with a cross-Kerr coupling $\chi_L=\chi_{LH} f(g/\Delta)^2$ and another more mechanical, with a cross-Kerr coupling $\chi_m = \chi_{LH} h(g/\Delta)^2$.
This gives rise to a HF spectrum
\begin{equation}
	\hbar\omega_{t,H}'-\chi_Ln_L-\chi_mn_m\ .
\end{equation}
For example, with $g/\Delta = 1/4, 1/2, 1$, we obtain $\chi_m/\chi_L \simeq 0.06,0.17,0.38$.
Different values allow the resolution of more mechanical peaks between two electrical peaks, and the optimum will depend on the specifics of an experiment.
To conclude this section, resolution of the first mechanical Fock states is possible for $g/\Delta\sim 1$ such that $\chi_m = \chi_{LH} /3$ and $\chi_L = 2\chi_{LH} /3$ if
\begin{equation}
	\gamma_{H,\text{eff}} \ll \chi_{LH} /3\hbar\ ,
\end{equation}
where $\gamma_{H,\text{eff}}$ is the effective line-width of the HF mode.
An example of the obtainable spectrum is given in Fig.\ref{fig:A_chi_g}.

\subsection{Case \texorpdfstring{$\chi_{LH}\gg \hbar g\gg A_L$}{chi>>g>>A}}

\begin{figure}
\centering
\includegraphics[width=0.75\textwidth]{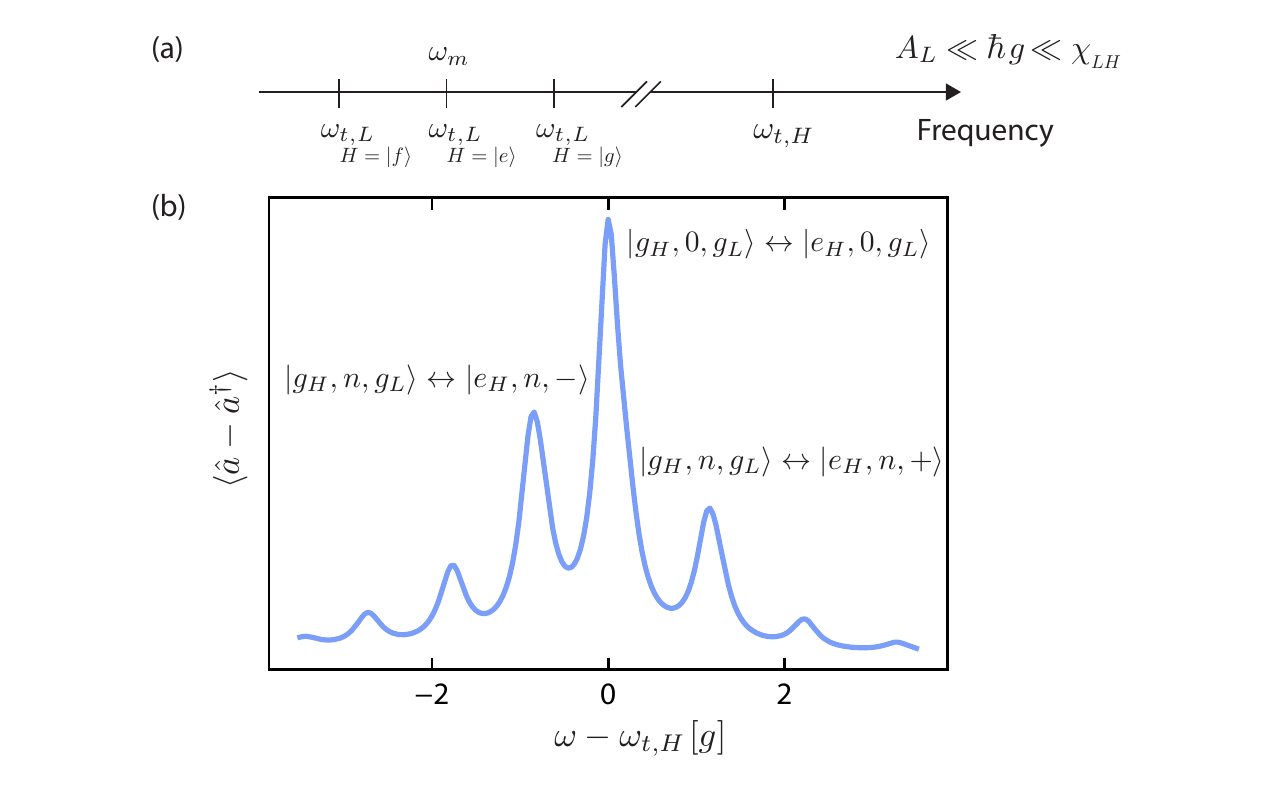}
\caption{
\textbf{HF mode spectrum in the case }$A_L\ll \hbar g\ll \chi_{LH}$.
{(a) Frequency landscape.}
With dominating $\chi_{LH}$, the LF mode frequency is a function of the state of the high frequency (HF) mode.
We focus on the case where the mechanical oscillator is resonant with the LF mode if the HF mode is in state $\ket{e}$.
{(b) Spectrum.} 
We plot the expectation value $\langle\hat a - \hat a^\dagger\rangle$ whilst weakly driving the system at a frequency $\omega$.
This corresponds to one quadrature in a homodyne measurement of the HF mode.
With the condition $\gamma_{t,H} + 4n_{\text{th}}\gamma_{t,L}\ll g$, with $\gamma_{H}$,$\gamma_{L}$ the dissipation rates of the HF and LF mode respectively, we can resolve phonon-number dependent transitions at frequencies $\omega_t \pm ng$, where $n$ is the number of phonons in the mechanical oscillator.
The exact simulation parameters and numerical methods are provided in Sec.~\ref{sec:numerical_methods}
}
\label{fig:A_g_chi}
\end{figure}

In this limit, the cross-Kerr should impose the relelevant basis, and the coupling be treated as a perturbation only.
We write the Hamiltonian as
\begin{equation}
\begin{split}
	\hat H =& \sum_j\ket{j}\bra{j}\otimes\left[E_j\right.\\
	+&  (\hbar\omega_{t,L}'-j\chi_{LH}) \hat b^\dagger \hat b +  (\hbar\omega_{t,L}'-\chi_{LH}) \hat c ^\dagger \hat c\\
	-& \hbar g(\hat b ^\dagger \hat c+\hat b  \hat c^\dagger)\left.\right]\ ,
\end{split}
\end{equation}
where $j$ denotes the state of the HF mode.
What is emphasized here, is that for $\chi_{LH}\gg \hbar g$, the two LF modes will only couple for certain values of $j$.
We explore the case where the mechanical mode is resonant with the LF electrical mode when the HF mode is in its first excited state.
The case where the two low frequency modes are resonant for the HF mode in its ground state yields similar results, the advantage here is that the HF spectrum reflects the occupation of the un-coupled mechanical and LF mode, rather than a hybridized one.
We now apply the unitary transformation $\sum_j\ket{j}\bra{j}\hat U_j$, where the unitary transformation $\hat U_j$ acts upon the hilbert space of the two low frequency electromechanical modes.
The transformation $\hat U_j$ should be the ones which move the two coupled frequency electromechanical modes to a new normal-mode basis (one for each state of the HF mode $j$), with annihilation operators and frequencies defined as in Eqs.~(\ref{eq:mode_1_annihilation},\ref{eq:mode_1_freqs}).
For $j=0$ and $j\le 2$ the two modes are off-resonant by at least $ \chi_{LH} \gg \hbar g$, such that we can apply the approximation $g\ll\Delta$ in the Eqs.~(\ref{eq:mode_1_annihilation},\ref{eq:mode_1_freqs}) leading to un-altered modes $\hat b_j = \hat b$ and $\hat c_j = \hat c$.
For $j=1$, the two modes are near resonance, leading to two normal modes $\hat \beta_\pm$ defined by
\begin{equation}
\hat \beta_\pm = (\hat b \pm \hat c)/\sqrt{2}\ ,
\end{equation}
and mode frequencies $\omega_{t,L}'-\chi_{LH}/\hbar\pm g$.
The Hamiltonian becomes
\begin{equation}
\begin{split}
	\hat H = \ket{g}\bra{g}\otimes&\left[\hbar \omega_{t,L}' \hat b^\dagger \hat b +  (\hbar\omega_{t,L}'-\chi_{LH}+\hbar\Delta) \hat c ^\dagger \hat c\right]\\
	+ \ket{e}\bra{e}\otimes&\left[\hbar\omega_{t,H}' +  (\hbar\omega_{t,L}'-\chi_{LH} - \hbar g) \hat \beta_-^\dagger \hat \beta_-+  (\hbar\omega_{t,L}'-\chi_{LH} + \hbar g) \hat \beta_+ ^\dagger \hat \beta_+\right]+...\\
\end{split}
\end{equation}
Probing the spectrum around $\omega_{t,H}'$ will reveal peaks at the following frequencies
\begin{align}
\begin{split}
&(\omega_{t,H}' + n_L(\omega_{t,L}'-\chi_{LH}/\hbar-g)+n_m(\omega_{t,L}'-\chi_{LH}/\hbar+g))-( n_L\omega_{t,L}'+n_m(\omega_{t,L}'-\chi_{LH}/\hbar))\ ,\\
&(\omega_{t,H}' + n_m(\omega_{t,L}'-\chi_{LH}/\hbar-g)+n_L(\omega_{t,L}'-\chi_{LH}/\hbar+g))-(n_L\omega_{t,L}'+n_m(\omega_{t,L}'-\chi_{LH}/\hbar))\ ,
\label{eq:chapte-5_RFcQED_spectrum_small_g}
\end{split}
\end{align}
which couple the only states with some overlap
\begin{align}
\begin{split}
\ket{g,n_L,n_m}&\leftrightarrow\ket{e,n_{-}=n_L,n_{+}=n_m}\ ,\\
\ket{g,n_L,n_m}&\leftrightarrow\ket{e,n_{-}=n_m,n_{+}=n_L}
\end{split}
\end{align}
respectively.
Here the eigenstates $\ket{g,n_L,n_m}$ correspond the the HF mode in the ground state and the LF and mechanical modes populated with $n_L$,$n_m$ photons or phonons respectively.
The eigenstates $\ket{e,n_+,n_-}$ correspond the the HF mode in the excited state and the \textit{hybridized} low frequency electro-mechanical modes populated with $n_+$,$n_-$ excitations respectively.
Eq.~(\ref{eq:chapte-5_RFcQED_spectrum_small_g}) can be re-written as
\begin{align}
\begin{split}
&\omega_{t,H}' - n_L(\chi_{LH}/\hbar-g)-n_mg\ ,\\
&\omega_{t,H}' - n_L(\chi_{LH}/\hbar+g)+n_mg\  .
\end{split}
\end{align}
In this regime, the HF mode is mostly sensitive to the LF mode, and detection of mechanical Fock states necessitates
\begin{equation}
	\gamma_{H,\text{eff}}\ll g \ll \chi_{LH}/\hbar\ ,
\end{equation}
where $\gamma_{H,\text{eff}}$ is the effective line-width of the HF mode.
An example of the obtainable spectrum is given in Fig.\ref{fig:A_g_chi}.

\subsection{Effective HF mode line-width \texorpdfstring{$\gamma_{H,\text{eff}}$}{gamma H} and conclusion}.
In order to determine an expression for $\gamma_{H,\text{eff}}$, we make the assumption that the contribution coming from the coupling to the lower frequency electro-mechanical modes is dominated by the line-width of the electrical losses.
By extending Eq.~(S26) of Ref.~\cite{gely2019observation}, we then have
\begin{equation}
	\gamma_{H,\text{eff}} = \gamma_{t,H} + 2\gamma_{t,L}(\langle n_L\rangle+(1+2\langle n_L\rangle)n_{\text{th}}) \ .
\end{equation}
Where $\langle n_L\rangle$ measures the average occupation of the states involved in a transition of interest, not the average number of photons populating the low frequency mode.
In order to derive a compact set of requirements valid for all above cases, we make the assumption that a representative measure of $\gamma_{H,\text{eff}}$ corresponds to transitions between states with $\langle n_L\rangle \sim 1/2$ \textit{i.e.} mechanical Fock states which are half hybridized with the LF mode.
We also assume that the thermal occupation of the low frequency modes is significant $n_{\text{th}}\gg1$, yielding 
\begin{equation}
	\gamma_{H,\text{eff}} \simeq \gamma_{t,H} + 4n_{\text{th}}\gamma_{t,L}\ .
\end{equation}
This effective linewidth can be injected into the requirements for phonon-number resolution derived in the four previous subsections
\begin{equation}
\begin{split}
	&\gamma_{H,\text{eff}}\ll\chi_{LH}/15\hbar\ ,\\
	&\gamma_{H,\text{eff}}\ll g/2\ ,\\
	&\gamma_{H,\text{eff}} \ll \chi_{LH} /3\hbar ,\\
	&\gamma_{H,\text{eff}}\ll g \ll \chi_{LH}/\hbar\ ,
\end{split}
\end{equation}
to produce the approximate requirement quoted in Eq.~(\ref{eq:main_ineq_3body}) which summarizes the results of this section
\begin{equation}
		\gamma_{t,H}+4\gamma_{t,L} n_{\text{th}}\ll\chi_{LH}/\hbar,g\ll\omega_m',\omega_{t,L}'\ .
\end{equation}

\section{Numerical methods}
\label{sec:numerical_methods}

Here we explain the methods used to generate Figs.~\ref{fig:g_A},\ref{fig:A_g},\ref{fig:g_A_chi},\ref{fig:A_g_chi},\ref{fig:chi_g_A},\ref{fig:A_chi_g}.
We aim to emulate the spectrum that one would measure experimentally from the previously derived Hamiltonians.
In the case where the mechanical oscillator is only coupled to a single transmon, we consider probing the transmon by adding a weak drive term 
\begin{equation}
	H_\text{dr}(t) = -i\hbar\gamma\times10^{-3}(\hat ae^{i\omega_d}-\hat a^\dagger e^{-i\omega_d})\ ,
\end{equation}
where $\omega_d$ is the driving frequency, and our choice of drive strength $\gamma\times10^{-3}$ does not significantly populate the driven mode.
We then move to the rotating frame of the drive through the unitary transformation
\begin{equation}
	\hat U = e^{i\hbar\omega_d(\hat a^\dagger\hat a+\hat c^\dagger\hat c)}\ ,
\end{equation}
such that the transformed Hamiltonian, including the coupling term, remains time-independent
\begin{equation}
	\hat H+H_\text{dr}(t)\rightarrow \hat U^\dagger \hat H \hat U = H +H_\text{dr}(0)-\hbar\omega_d(\hat a^\dagger\hat a+\hat c^\dagger\hat c)\ .
\end{equation}
We follow a similar process in the case where two electrical modes are considered, where we wish to probe the high-frequency mode.
In this case the unitary transformation is 
\begin{equation}
	\hat U = e^{i\hbar\omega_d\hat a^\dagger\hat a}\ ,
\end{equation}
\begin{equation}
	\hat H+H_\text{dr}(t)\rightarrow \hat U^\dagger \hat H \hat U = H +H_\text{dr}(0)-\hbar\omega_d\hat a^\dagger\hat a\ .
\end{equation}
We make use of QuTiP~\cite{johansson2012qutip,johansson2013qutip} to compute the steady-state using a Lindblad master equation, where each mode is subjected an interaction with the environment characterized by two collapse operators.
For example, for the mechanical mode these collapse operators write
\begin{equation}
	\sqrt{\gamma_m(1+n_\text{th})}\ \hat c\ \text{and}\ \sqrt{\gamma_mn_\text{th}}\ \hat c^\dagger\ .
\end{equation}
We then use the steady-state density matrix to compute and plot the expectation value of $\hat a + \hat a^\dagger$ to emulate a homodyne measurement of the systems spectrum.
The parameters used in the simulation are summarized in the table~\ref{tab:simulation_parameters}.
The code used to generate the the associated figures is available in Zenodo with the DOI identifier 10.5281/zenodo.4292367 .

\begin{table}[h!]
	{\renewcommand{\arraystretch}{1.2}
    \begin{tabular} {|c|cccc|c|ccccc|c|ccccc|}
        \cline{2-17}
        \multicolumn{1}{c|}{}&\multicolumn{4}{c|}{mechanical oscillator}&$\leftrightarrow$&\multicolumn{5}{c|}{LF electrical mode}&$\leftrightarrow$&\multicolumn{5}{c|}{HF electrical mode}\\
        \cline{2-17}
        \multicolumn{1}{c|}{}&$\omega$&$\gamma$&$n_\text{th}$&$N$&$g$&$\omega$&$A$&$\gamma$&$n_\text{th}$&$N$&$\chi_{LH}$&$\omega$&$A$&$\gamma$&$n_\text{th}$&$N$ \\ \hline
        Fig.~\ref{fig:g_A}&$1$&$10^{-7}$&$1.2$&$6$&$0.005$&$1$&$0.05$&$10^{-4}$&$1.2$&$4$&&&&&& \\ \hline
        Fig.~\ref{fig:A_g}&$1$&$10^{-7}$&$1.2$&$6$&$0.75$&$1$&$0.005$&$4\times10^{-5}$&$1.2$&$6$&&&&&& \\ \hline
        Fig.~\ref{fig:chi_g_A}&$1.0078$&$5\times10^{-7}$&$0.5$&$4$&$5\times10^{-3}$&$1$&$0.05$&$5\times10^{-7}$&$0.5$&$4$&$5\times10^{-4}$&$50$&$2.5$&$5\times10^{-6}$&$0$&$3$ \\ \hline
        Fig.~\ref{fig:g_A_chi}&$1$&$10^{-6}$&$0.5$&$4$&$5\times10^{-3}$&$1$&$0.05$&$10^{-6}$&$0.5$&$4$&$5\times10^{-4}$&$50$&$2.5$&$10^{-4}$&$0$&$3$ \\ \hline
        Fig.~\ref{fig:A_chi_g}&$1.2$&$10^{-5}$&$0.5$&$4$&$0.1$&$1$&$0.001$&$10^{-5}$&$0.5$&$4$&$0.01$&$50$&$0.025$&$10^{-4}$&$0$&$3$ \\ \hline
        Fig.~\ref{fig:A_g_chi}&$85\times10^6$&$850$&$3.68$&$4$&$2\times10^6$&$10^8$&$225\times10^3$&$10\times10^3$&$3.68$&$4$&$15\times10^6$&$5\times10^9$&$250\times10^6$&$0.5\times10^6$&$0$&$3$ \\ \hline
    \end{tabular} \\ \vskip .5cm
	}
    \caption{System parameters used to generate Figs.~\ref{fig:g_A},\ref{fig:A_g},\ref{fig:g_A_chi},\ref{fig:A_g_chi},\ref{fig:chi_g_A},\ref{fig:A_chi_g}. $\omega$,$A$,$\gamma$,$n_\text{th}$ and $N$ designate the frequency, anharmonicity, dissipation rate, thermal occupancy and Hilbert space size of the different modes of the system respectively. The coupling between the low-frequency electrical mode and the mechanical oscillator is denoted by $g$ and the cross-Kerr interaction between the two electrical modes $\chi_{LH}$.}
    \label{tab:simulation_parameters}   
\end{table}

\bibliography{library}

\end{document}